%% file: amain.tex
\documentclass{article}

\usepackage{mathtools}
\usepackage{setspace}
\usepackage{amsmath}
\usepackage{amsthm}
\usepackage{amssymb}
\usepackage{titlesec}
\usepackage{mymacros}
\usepackage{xspace}
\usepackage{todonotes}
\usepackage[margin=1.25in]{geometry}
\usepackage{subfiles}

\newtheorem{thm}{Theorem}[section]
\newtheorem{theorem}[thm]{Theorem}
\newtheorem{claim}[thm]{Claim}

\newtheorem{definition}[thm]{Definition}

\newtheorem{lemma}[thm]{Lemma}
\newtheorem{prop}[thm]{Proposition}
\newtheorem{cor}[thm]{Corollary}

\newtheorem{example}[thm]{Example}

\DeclareMathOperator{\id}{id}

\begin{document}

\title{Lower Bounds for Symmetric Circuits for the Determinant\thanks{Research funded by EPSRC
    grant EP/S03238X/1.  A preliminary version of this paper appeared as~\cite{DW-itcs}}}

\author{Anuj Dawar and Gregory Wilsenach \\
  Department of Computer Science and Technology\\ University of Cambridge.\\
  \texttt{anuj.dawar@cl.cam.ac.uk, gregory.wilsenach@cl.cam.ac.uk}}

\maketitle

\begin{abstract}
  Dawar and Wilsenach (ICALP 2020) introduce the model of symmetric arithmetic circuits and show an exponential separation between the sizes of symmetric circuits for computing the determinant and the permanent.  The symmetry restriction is that the circuits which take a matrix input are unchanged by a permutation applied simultaneously to the rows and columns of the matrix.  Under such restrictions we have polynomial-size circuits for computing the determinant but no subexponential size circuits for the permanent.  Here, we consider a more stringent symmetry requirement, namely that the circuits are unchanged by arbitrary even permutations applied separately to rows and columns, and prove an exponential lower bound even for circuits computing the determinant.  The result requires substantial new machinery.  We develop a general framework for proving lower bounds for symmetric circuits with restricted symmetries, based on a new support theorem and new two-player restricted bijection games.  These are applied to the determinant problem with a novel construction of matrices that are bi-adjacency matrices of graphs based on the CFI construction.  Our general framework opens the way to exploring a variety of symmetry restrictions and studying trade-offs between symmetry and other resources used by arithmetic circuits.
\end{abstract}

\section{Introduction}\label{sec:introduction}

\subfile{introduction}

\section{Background}\label{sec:background}

\subfile{background}

\section{Symmetric Circuits}\label{sec:symm-circ}
\subfile{symmetric-circuits}

\section{Games and Supports}\label{sec:games-supports}
\subfile{games-and-supports}

\subsection{Bijection Games on Indexed function}\label{sec:bij-games}

\subfile{new-bijection}

\subsection{Supports}\label{sec:supports}

\subfile{supports}

\subsection{Playing Games on Circuits}\label{sec:games-circuits}

\subfile{games-on-circuits}

\subsection{Bounds on Supports}\label{sec:bounds-supports}

\subfile{bounds-on-supports}

\section{Lower Bound for the Determinant}\label{sec:lower-bound}

\subfile{lower-bound}

\subsection{Graphs with Odd Number of Perfect
  Matchings}\label{sec:base-graph}
\subfile{base-graph}

\subsection{Playing the Game}\label{sec:game}

\subfile{playing}

\subsection{Bringing it Together}\label{sec:main-proof}
\subfile{main-proof}

\section{Lower Bound for the Permanent}\label{sec:permanent}

\subfile{permanent}

\section{Concluding Discussion}

\subfile{conclusions}

\bibliographystyle{plain} \bibliography{references.bib}
\end{document}

%% file: introduction.tex
The central open question in the field of arithmetic circuit complexity is the
separation of the complexity classes $\VP$ and $\VNP$. Sometimes known as
Valiant's conjecture, this is also described as the algebraic analogue of the
$\PT$ vs.\ $\NP$ question. The conjecture is equivalent to the statement that
the permanent of a matrix cannot be expressed by a family of polynomial-size
arithmetic circuits. Lower bounds on the size of circuits computing the
permanent have been established by imposing certain restrictions on the circuit
model. For instance, it is known that there is no subexponential family of \emph{monotone} circuits for the permanent~\cite{JerrumS82} and an
exponential lower bound for the permanent is also known for \emph{depth-3}
arithmetic circuits~\cite{GK98}.  In both these cases, the lower bound obtained
for the permanent also applies to the determinant, which is known to be in
$\VP$\footnote{Note that the determinant is not itself monotone in the usual
  sense, but there is a suitably adapted notion of monotonicity called
  \emph{syntactic monotonicity} in~\cite{KayalS14} with respect to which the determinant does have circuits, but not subexponential size ones.}.
In that sense, the lower bounds tell us more about the weakness of the
model resulting from the restriction than the difficulty of computing the
permanent.

In this paper we focus on another restriction on arithmetic circuits introduced
relatively recently: that of symmetry~\cite{DawarW20}. This has been shown to
give an exponential separation in the size of circuits computing the permanent
and the determinant. We first introduce this restriction informally (a more
formal definition is given in Section~\ref{sec:symm-circ}). Given a field $\ff$
and a set of variables $X$, let $C$ be a circuit computing a polynomial $p$ in
$\ff[X]$. For a group $G$ acting on the set $X$, we say that $C$ is
$G$-symmetric if the action of any element $g \in G$ on the inputs of
$C$ can be extended to an
automorphism of $C$. Of course, this makes sense only when the polynomial $p$
itself is invariant under the action of $G$.

For example, both the permanent and the determinant are polynomials in a
\emph{matrix} of variables $X = \{ x_{ij} \mid 1 \leq i,j \leq n\}$. Let $G$ be
the group $\sym_n$ acting on $X$ by the action whereby $\pi \in G$ takes
$x_{ij}$ to $x_{\pi(i)\pi(j)}$. We call this the \emph{square symmetric} action.
It corresponds to arbitrary permutations applied simultaneously to the rows and
columns of the matrix. We show in~\cite{DawarW20} that there are polynomial-size
$G$-symmetric circuits computing the determinant, but any family of
$G$-symmetric circuits for computing the permanent has exponential size. Both
results are established for any field of characteristic zero.

The choice of the group action $G$ in these results is natural, but certainly
not the only possibility. The lower bound immediately applies to any larger
group of symmetries of the polynomial as well. Consider, for instance the action
of the group $\sym_n \times \sym_n$ on $X$ whereby $(\pi,\sigma)$ takes
$x_{ij}$ to $x_{\pi(i)\sigma(j)}$. We call this the \emph{matrix symmetric}
action. It corresponds to independent permutations applied to the rows and
columns. The permanent is invariant under this action and the exponential lower
bound for square-symmetric circuits for the permanent applies \emph{a fortiori}
to matrix-symmetric circuits as well. As it happens, in this case we can
strengthen the lower bound, if not in terms of size at least in terms of the
range of application. That is, the exponential lower bound has been shown not
only for circuits computing the permanent in fields of characteristic zero, but
over all fields of characteristic other than two. This exponential lower bound
matches known upper bounds as the smallest circuits for computing permanents,
those based on Ryser's formula, are in fact $\sym_n \times \sym_n$-symmetric.

In the case of the determinant, it was left open whether the polynomial upper
bound obtained for square symmetric circuits could be improved by requiring
larger groups of symmetries on the circuits. The most efficient algorithms for
computing the determinant (based on Gaussian elimination) are not square symmetric and the polynomial upper bound is obtained by an application
of Le Verrier's method~\cite{DawarW20}. It is a natural question to ask how much
more stringent a symmetry requirement we can impose and still find efficient
algorithms. The determinant is not matrix-symmetric like the permanent is. Let
us write $\mathbf{D}_n$ for the group of permutations of $X = \{ x_{ij} \mid 1
\leq i,j \leq n\}$ which fix the determinant. This can be seen to be $D_n \ltimes
T$ where $D_n$ is the subgroup of $\sym_n \times \sym_n$ of index $2$ consisting
of pairs $(\sigma,\pi)$ of permutations with $\sgn{\sigma} = \sgn{\pi}$ and $T
\cong \mathbb{Z}_2$ represents the transposition of rows and columns. We prove
in the present paper that any family of $\mathbf{D}_n$-symmetric circuits
computing the determinant must have exponential size. Indeed, our lower bound is
proved even for the subgroup of $D_n$ given by $\alt_n\times \alt_n$.

Proving this lower bound requires substantially different methods than those of
other results, and developing these methods is a central thrust of this paper.
The exponential lower bound for square-symmetric circuits for the permanent is
established in~\cite{DawarW20} by proving a lower bound on the \emph{orbit size}
of Boolean circuits computing the permanent of a $\{0,1\}$-matrix. Here, the
orbit size of a circuit $C$ is the maximal size of an orbit of a gate of $C$
under the action of the automorphism group of $C$. This lower bound is proved
using a connection between the orbit size of circuits computing a graph
parameter and the \emph{counting width} of the parameter as established
in~\cite{AndersonD17}. To be precise, it is shown that if a graph parameter has
linear counting width, i.e.\ it distinguishes graphs on $n$ vertices which are
not distinguished by $\Omega(n)$-dimensional Weisfeiler-Leman equivalences, then
it cannot be computed by symmetric circuits of subexponential orbit size. The
Weisfeiler-Leman equivalences are well-studied approximations of the graph
isomorphism relation, graded by dimension (see~\cite[Section~3.5]{grohe2017descriptive}
for an introduction). The equivalences have many equivalent characterizations
arising in combinatorics, algebra, logic and linear optimization. The term
counting width comes from the connection with counting logic
(see~\cite{Dawar-CSL}). The main technical ingredient in the lower bound proof
in~\cite{DawarW20} is then a proof of a linear lower bound on the counting width
of the number of perfect matchings in a bipartite graph.

We were able to rely on lower bounds on the counting width of graph parameters
because a $\sym_n$-invariant parameter of a $\{0,1\}$-matrix can be seen as a
graph parameter. That is, a graph parameter that does not distinguish between
isomorphic graphs is necessarily $\sym_n$-invariant on the adjacency matrices of
graphs. Similarly, the $\sym_n \times \sym_n$ action on a matrix can be
understood as the natural invariance condition of the biadjacency matrix of a
bipartite graph. On the other hand, there seems to be no natural graph structure
giving rise to an $\alt_n\times \alt_n$-invariance requirement on the matrices.
For this reason, we develop here both a general framework for presenting and
studying symmetric circuits and new methods for proving lower bounds under some
of these symmetry assumptions.

The generality of this framework allows us to consider a variety of different
symmetry conditions, providing both a broad vocabulary for working with these
circuits and game-based characterizations of the expressive power of these
various symmetric models. This opens up the possibility of studying symmetry as
a resource. Our results suggest a spectrum of symmetry restrictions and it would
be interesting to establish exactly where on this spectrum the boundary of
efficient algorithms for the determinant lies. Similar questions can be asked
about other polynomials which admit efficient evaluation. For the permanent, the
natural question is how much can we relax the symmetry conditions and still
prove lower bounds. This is all to say that, quite apart from the main result,
we regard the framework developed here as a contribution in its own right, which
lays out a landscape to explore symmetry as a resource in this area.

Table~\ref{tab:results} summarizes what is currently known about the power of
various symmetric circuit models computing the permanent and determinant. The
first column is for unrestricted circuits, i.e.\ those symmetric under the
trivial group action. The upper bound for the determinant is by an adapted
Gaussian elimination algorithm~\cite{Burgisser1997} and the upper bound for the
permanent, which in fact holds for every group in the table, is by Ryser's formula~\cite{Ryser1963}. The lower bound in the first
column is the trivial one. The second and fourth column state results
established in~\cite{DawarW20}. There is no result for the determinant in the
fourth column as the determinant is not invariant under the $\sym_n \times
\sym_n$ action. The results in the column for $\alt_n \times \alt_n$ are new to
this paper. In the last two columns $\mathbf{D}_n$ and $\mathbf{P}_n$ represent
the full invariance groups (as subgroups of $\sym_{n\times n}$) of the
determinant and permanent respectively. The lower bounds stated in those columns
follow from the ones obtained for their subgroups $\alt_n \times \alt_n$ and
$\sym_n \times \sym_n$ respectively.

A more detailed discussion of some of the main innovations follows.
\begin{table}
  \centering
  \begin{tabular}{c | c | c | c | c | c | c |}
    $G$  &  $\{ \text{id} \}$ & $\sym_n$ & $\alt_{n} \times \alt_{n}$ &  $\sym_{n} \times \sym_{n}$ & $\mathbf{D}_n$ & $\mathbf{P}_n$ \\ \hline & & & & & & \\
    {Det} &  $O(n^{3}) $ & \begin{tabular}{c}
                             $O(n^4)$ \\ \emph{(char 0)}
                           \end{tabular}
         &   \begin{tabular}{c}       $2^{\Omega(n)}$ \\ \emph{(char 0)}
             \end{tabular} & {N/A}
                              &   \begin{tabular}{c}       $2^{\Omega(n)}$ \\ \emph{(char 0)}
                                  \end{tabular} & {N/A}
    \\ & & & & & & \\ \hline   & & & & & &  \\
    Perm &
           \begin{tabular}{c}
             $O(n^22^n)$ \\  $\Omega(n^2)$
           \end{tabular}
         & \begin{tabular}{c}  $2^{\Omega(n)}$ \\ \emph{(char 0)}
           \end{tabular}
         & \begin{tabular}{c} $2^{\Omega(n)}$ \\ \emph{(char $\neq 2$)} \end{tabular}
         & \begin{tabular}{c}
             $2^{\Omega(n)}$ \\ \emph{(char $\neq 2$)} \end{tabular}
         & \begin{tabular}{c} $2^{\Omega(n)}$ \\ \emph{(char $\neq 2$)} \end{tabular}
         & \begin{tabular}{c}
             $2^{\Omega(n)}$ \\ \emph{(char $\neq 2$)} \end{tabular}
     \\  & & & & & &
  \end{tabular}
  \caption{Table of upper and lower bounds for $G$-symmetric circuits for the
    determinant and permanent, for various group actions $G$. Here
    $\{\text{id}\}$  denotes the trivial group, and thus the column
    gives upper and lower bound for general circuits.}
  \label{tab:results}
\end{table}

\begin{enumerate}
\item The lower bounds on counting width of graph parameters are proved using
  the method of \emph{bijection games} applied to graphs based on a construction
  due to Cai, F\"{u}rer and Immerman~\cite{CFI92} (we refer to this class of
  graph constructions as the \emph{CFI construction}). The games are
  parameterized by a natural number $k$ and are used to establish the
  indistinguishability of graphs by means of $k$-dimensional Weisfeiler-Leman
  equivalences. We adapt the bijection games and show that they can be directly
  used to obtain lower bounds for symmetric circuits without reference to graphs
  or Weisfeiler-Leman equivalences. The parameter $k$ is related to a parameter
  of the circuits we call \emph{support size}. This is based on an idea outlined
  in~\cite{Dawar2016} which we elaborate further.
\item The support theorem established in~\cite{AndersonD17} and stengthened
  in~\cite{DawarW20} is a key tool, which we extend further. It shows that small
  symmetric circuits have small support size. This was proved for symmetry under
  the action of the symmetric group originally and we extend the range of group
  actions for which it can be shown to hold.
\item The original $k$-pebble bijection games of Hella~\cite{Hella1996} are
  two-player games played on a pair of graphs (or more generally relational
  structures) between two players called \emph{Spoiler} and \emph{Duplicator}.
  In each move of the game, Duplicator is required to provide a bijection
  between the two graphs.  A winning strategy for Duplicator shows that the two
  graphs are not distinguishable in the $k$-dimensional Weisfeiler-Leman
  equivalences.  We refine the games by restricting Duplicator to play bijections
  from a restricted set of bijections.  This set is obtained by
  composing an initial bijection with permutations from some group $G$. In this game,
  even an isormophism does not guarantee a Duplicator winning strategy. But, we
  are able to relate Duplicator winning strategies to indistinguishability by
  $G$-symmetric circuits taking the adjacency matrices as input.  At
  the same time, we generalize the game so it can be played on a much more general notion of structured input  rather than a relational structure.
\item A fourth key ingredient is the construction of matrices with distinct
  determinants on which we can show Duplicator winning strategies in the $\alt_n
  \times \alt_n$-restricted bijection game. The two matrices are obtained as
  biadjacency matrices of a single bipartite graph given by the CFI
  construction. The two matrices differ only through the interchange of two
  columns. The challenge here is to show that they have non-zero determinant.
\item An important element of the construction of the matrices in (4) are
  bipartite $3$-regular graphs with large tree-width and an odd number of
  perfect matchings. We show the existence of an infinite family of such graphs
  in Section~\ref{sec:base-graph}. This may be of independent interest.  A
  strengthening of the conditions on the number of perfect matchings could be
  used to extend our lower bound to fields of positive characteristic other than
  two. We expand on this in Section~\ref{sec:main-proof}.
\end{enumerate}

Our work may be compared to that of Landsberg and Ressayre~\cite{LandsbergR16}
who establish an exponential lower bound on the complexity of the permanent
(specifically over the complex field $\mathbb{C}$) under an assumption of symmetry. Their
lower bound is for equivariant determinantal representations of the permanent,
that is those that preserve all the symmetries of the permanent function. This
approach doesn't yield any lower bounds for symmetric circuits in the sense we
consider here.  On the other hand, lower bounds for equivariant
determinantal complexity can be derived from the results
in~\cite{DawarW20}, albeit not ones as strong as
in~\cite{LandsbergR16}.  For a more detailed comparison of the two
approaches see the full version of~\cite{DawarW20}.

This paper is organized as follows. We begin with introducing the notation we need in Section~\ref{sec:background}.
Symmetric circuits working on arbitrary structured inputs are defined in
Section~\ref{sec:symm-circ}.  The support theorem we need is proved in
Section~\ref{sec:supports}.  The bijection games are defined in
Section~\ref{sec:bij-games} where we also prove that Duplicator winning
strategies imply indistinguishability by small circuits.  The main construction
establishing the lower bound for the determinant is presented in
Section~\ref{sec:lower-bound}.  The corresponding lower bound for the permanent
is given in Section~\ref{sec:permanent}.  The proof for the last result is only
sketched as we only need to note that the proof from~\cite{DawarW20} can be
easily adapted to the new games we define to tighten the symmetry result from
$\sym_n \times \sym_n$ to $\alt_n \times \alt_n$.

%% file: background.tex
In this section we discuss relevant background and introduce notation.

We write $\nats$ for the positive integers and $\natz$ for the non-negative
integers. For $m \in \natz$, $[m]$ denotes the set $\{1, \ldots, m \}$. For a
set $X$ we write $\pow(X)$ to denote the powerset of $X$. We write $\id$ to
denote the identity function on some specified set. For $f : X \ra Y$ and $S
\subseteq X$ we write $f(S)$ to denote the image of $S$. Let $\bij(A, B)$ denote
the set of bijections from $A$ to $B$.

\subsection{Groups}
Let $\sym_A$ and $\alt_A$ denote the symmetric group and alternating group on
the set $A$. We write $\sym_n$ and $\alt_n$ to abbreviate $\sym_{[n]}$ and
$\alt_{[n]}$, respectively. Let $\{\id\}$ denote the trivial group. For groups
$G$ and $H$ we write $H \leq G$ to denote that $H$ is a subgroup of $G$. The
\emph{sign} of a permutation $\sigma \in \sym_A$ is defined so that if $\sigma$
is even $\sgn{\sigma} = 1$ and otherwise $\sgn{\sigma} = -1$.


Let $G$ be a group acting on a set $X$. We denote this as a left action, i.e.\
$\sigma x$ for $\sigma \in G$, $x \in X$. The action extends in a natural way to
powers of $X$. So, for $(x,y) \in X \times X$, $\sigma(x,y) = (\sigma x,\sigma
y)$. It also extends to the powerset of $X$ and functions on $X$ as follows. The
action of $G$ on $\pow(X)$ is defined for $\sigma \in G$ and $S \in \pow(X)$ by
$\sigma S = \{\sigma x \mid x \in S\}$. For $Y$ any set, the action of $G$ on
$Y^X$ is defined for $\sigma \in G$ and $f\in Y^X$ by $(\sigma f) (x) = f(\sigma
x)$ for all $x \in X$. We refer to all of these as the \emph{natural action} of
$G$ on the relevant set.

A set $X$ with the action of group $G$ on it is called a $G$-set. We do not
distinguish notationally between a $G$-set $X$ and the underlying set of
elements. Thus, we can say that if $X$ is a $G$-set, the collection of functions
$Y^X$ is also a $G$-set with the natural action.

Let $X = \prod_{i \in I} X_i$ and for each $i \in I$ let $G_i$ be a group acting
on $X_i$. The action of the direct product $G := \prod_{i \in I}G_i$ on $X$ is
defined for $x = (x_i)_{i \in I} \in X$ and $\sigma = (\sigma_i)_{i \in I} \in
G$ by $\sigma x = (\sigma_i x_i)_{i \in I}$. If instead $X = \biguplus_{i \in
  I}X_i$ then the action of $G$ on $X$ is defined for $x \in X$ and $\sigma =
(\sigma_i)_{i \in I} \in G$ such that if $x \in X_i$ then $\sigma x = \sigma_i
x$. Again, we refer to either of these as the \emph{natural action} of $G$ on
$X$.

Let $X$ be a $G$-set. Let $S \subseteq X$. Let $\stab_G(S) := \{\sigma \in G \mid
\forall x \in S \,\, \sigma x = x\}$ denote the \emph{pointwise stabilizer} of
$S$. Let $\setstab_G(S) := \{\sigma \in G \mid \sigma (S) = S\}$ denote the
\emph{setwise stabilizer} of $S$. If $S = \{x\}$ is a singleton we omit set
braces and write $\stab_G(x)$. Note that $\setstab_G(S)$ is the pointwise stabilizer
of $\{S\}$ in the $G$-set $\pow(X)$. For $x \in X$ let $\orb_G(x) := \{\sigma (x)
\mid \sigma \in G\}$ denote the \emph{orbit} of $x$.  In all cases, we
omit the subscript $G$ where it is clear from context.

\subsection{Matrices}

Let $A$ and $B$ be finite non-empty sets. We identify matrices with rows indexed
by $A$ and columns by $B$ and entries from some set $X$ with functions of
the form $M : A \times B \ra X$. So for $a \in A$, $b \in B$, $M_{ab} = M(a,
b)$. We also denote $M$ by $(M_{ab})_{a \in A, b \in B}$, or just $(M_{ab})$
when the index sets are clear from context.

Let $R$ be a commutative ring and $M: A \times B \ra R$ be a matrix with $\vert
A \vert = \vert B \vert$. The \emph{permanent} of $M$ over $R$ is $\perm_R(M) =
\sum_{\sigma \in \bij(A, B)}\prod_{a \in A} M_{a \sigma(a)}$. Suppose $A = B$.
The \emph{determinant} of $M$ over $R$ is $\det_R (M) = \sum_{\sigma \in
  \sym_A}\sgn{\sigma}\prod_{a \in A} M_{a \sigma(a)}$. The \emph{trace} of $M$
over $R$ is $\trace_R(M) = \sum_{a \in A} M_{a a}$. We omit reference to the
ring when it is obvious from context.  When $R$ is a field, we write $\rank(M)$ to denote the \emph{rank} of the matrix $M$.

We always use $\ff$ to denote a field and $\chr(\ff)$ to denote the
characteristic of $\ff$. For any prime power $q$ we write $\ff_q$ for the finite
field of order $q$. We are often interested in polynomials and circuits defined
over a set of variables $X$ with a natural matrix structure, i.e.\ $X = \{x_{ab}
: a \in A, b \in B\}$. We identify $X$ with this matrix. We also identify any
function of the form $f : X \ra Y$ with the $A \times B$ matrix with entries in
$Y$ defined by replacing each $x_{ab}$ with $f(x_{ab})$.

\subsection{Graphs}
Given a graph $\Gamma = (V,E)$, the \emph{adjacency matrix} $A_\Gamma$ of
$\Gamma$ is the $V \times V$ $\{0,1\}$-matrix with $A_\Gamma(u,v) = 1$ if, and
only if, $\{u,v\} \in E$. If $\Gamma$ is bipartite, with bipartition $V = U \cup
W$, then the \emph{biadjacency matrix} $B_\Gamma$ of $\Gamma$ is the $U \times
W$ $\{0,1\}$-matrix with $B_\Gamma(u,v) = 1$ if, and only if, $\{u,v\} \in E$.
For a set $U \subseteq V$ we write $\Gamma[U]$ to denote the subgraph induced by $U$.

A \emph{$k$-factor} of a graph $\Gamma$ is a spanning $k$-regular subgraph. A
\emph{perfect matching} is a $1$-factor. It is well known that for a bipartite
graph $\Gamma$, $\perm(B_\Gamma)$ over any field of characteristic zero counts
the number of perfect matchings in $\Gamma$~\cite{Harary69} and for prime $p$,
$\perm_{\ff}(B_\Gamma)$ for a field $\ff$ of characteristic $p$ counts the
number of perfect matchings in $\Gamma$ modulo $p$.  For bipartite $\Gamma$,
$A_\Gamma$ is a block anti-diagonal matrix with two blocks corresponding to
$B_\Gamma$ and $B_\Gamma^T$, and so $\perm(A_\Gamma) = \perm(B_\Gamma)^2$.

Let $\Gamma = (V, E)$ be a graph. For $S \subseteq V$ let $N^+ (S) := \{v \in
V\setminus S \mid \exists s \in S, \,\, (v, s) \in E\}$. We say $\Gamma$ is an
\emph{$\alpha$-expander} for $\alpha \in [0, 1]$ if for every $S \subseteq V$ of
size at most $\vert V\vert /2$ we have $\vert N^+(S) \vert \geq \alpha \vert S
\vert$. For an introduction to expander graphs see~\cite{Hoory06}. A set $S$ of
vertices in a graph $\Gamma$ is a \emph{balanced separator} if no connected
component of $\Gamma\setminus S$ contains more than half the vertices of
$\Gamma$. It is easy to see that for any $\alpha$ there is a constant
$\tau$ such that if $\Gamma$ is an $\alpha$-expander then  $\Gamma$
has no balanced separator of size less than $\tau |V|$.

The tree width of a graph is a well-known graph parameter which can be
characterized in terms of the cops and robbers game~\cite{SEYMOUR199322}.
This game is played on a graph $\Gamma = (V, E)$, with one player controlling
the single robber and another a team of  $k$ cops. The cops and robbers both sit on
nodes in the graph. At the start of a round the cops are sitting on nodes $X
\subset V$ and the round begins with the cop player announcing he intends to
move cops from $X' \subseteq X$ to occupy the nodes $Y$. The robber player must
respond with a path from the robbers current location that does not intersect
$X \setminus X'$. At the end of the round the cops and robber are moved to their new
locations, and if any cop is on the same node as the robber the cops win. The
characterization is as follows.

\begin{thm}[\cite{SEYMOUR199322}]
  There is a winning strategy for the cop player with $k$ cops if, and only if,
  the tree-width of $\Gamma$ is at most $k-1$.
\end{thm}

\subsection{Circuits}
We give a general definition of a circuit that incorporates both Boolean and
arithmetic circuits.
\begin{definition}[Circuit]
  A \emph{circuit} over the \emph{basis} $\BB$ with \emph{variables} $X$ and
  \emph{values} $K$ is a directed acyclic graph with a labelling where each
  vertex of in-degree $0$ is labelled by an element of $X \cup K$ and each
  vertex of in-degree greater than $0$ is labelled by an element of $\BB$.
\end{definition}
Let $C$ be circuit that is the labelling of a directed graph $(D, W)$,
where $W \subset D \times D$ with values $K$.  We call the elements of
$D$ \emph{gates}, and the elements of $W$ \emph{wires}.  We call the
gates with in-degree $0$ \emph{input gates} and gates with out-degree
$0$ \emph{output gates}.  We call those input gates labelled by
elements of $K$ \emph{constant gates}.  We call those gates that are
not input gates \emph{internal gates}.  For $g, h \in D$ we say that
$h$ is a \emph{child} of $g$ if $(h, g) \in W$.  We write $\child{g}$
to denote the set of children of $g$.  We write $C_g$ to denote the
sub-circuit of $C$ rooted at $g$, i.e\ the sub-circuit induced by
those gates with a directed path to $g$.  Unless otherwise stated we always
assume a circuit has exactly one output gate.

If $K$ is a field $\ff$, and $\BB$ is the set $\{+, \times\}$, we have an
\emph{arithmetic circuit} over $\ff$. If $K = \{0,1\}$, and $\BB$ is a
collection of Boolean functions, we have a \emph{Boolean circuit} over the basis
$\BB$. We define two bases here. The first is the \emph{standard basis} $\BS$
containing the functions $\land$, $\lor$, and $\neg$. The second is the
\emph{threshold basis} $\BT$ which is the union of $\BS$ and $\{t_{\geq k} : k
\in \nats\}$, where for each $k \in \nats$, $t_{\geq k}$ is defined for a string
$\vec{x} \in \{0, 1\}^*$ so that $t_{\geq k}(\vec{x}) = 1$ if, and only if, the
number of $1$s in $\vec{x}$ at least $k$. We call a circuit defined over this
basis a \emph{threshold circuit}. Another useful Boolean function is $t_{=k}$,
which is defined by $t_{=k}(x) = t_{\leq k}(x) \land \neg t_{\leq k+1}(x)$. We
do not explicitly include it in the basis as it is easily defined in $\BT$.

In general, we require that a basis contain only functions that are invariant
under all permutations of their inputs. That is, if $f \in \BB$ is such that
$f: K^n \ra K$ for some $n \in \nats$ then for all $\sigma \in \sym_n$ and $M
\in K^n$ we have $f (M \circ \sigma) = f(M)$. We call such a function $f$
\emph{fully symmetric}. The arithmetic functions $+$ and $\times$ and all of
the Boolean functions in $\BT$ and $\BS$ are fully symmetric. Let $C$ be a
circuit defined over such a basis with variables $X$ and values $K$. We evaluate
$C$ for an assignment $M \in K^X$ by evaluating each gate labelled by some $x
\in X$ to $M(x)$ and each gate labelled by some $k \in K$ to $k$, and then
recursively evaluating each gate $g$ according to its corresponding
basis element.  That is $g$ gets the value obtained by applying the
function labelling $g$ to the set of values of $\child{g}$.
We write $C[M](g)$ to denote the value of the gate $g$ and $C[M]$ to denote the
value of the output gate. We say that $C$ computes the function $M \mapsto
C[M]$.

It is conventional to consider an arithmetic circuit $C$ over $\ff$ with
variables $X$ to be computing a polynomial in $\ff[X]$, rather than a function
$\ff^X \ra \ff$. This polynomial is defined via a similar recursive evaluation,
except that now each gate labelled by a variable evaluates to the corresponding
formal variable, and we treat addition and multiplication as ring operations in
$\ff[X]$. Each gate then evaluates to some polynomial in $\ff[X]$. The
polynomial computed by $C$ is the value of the output gate.

For more details on arithmetic circuits see~\cite{ShpilkaY10} and for Boolean
circuits see~\cite{Vollmer99}.

%% file: symmetric-circuits.tex
Complexity theory is often concerned with computation models that take as
input binary strings. In practice these strings are almost always taken to
encode some structured input (e.g.\ graphs, matrices, numbers, etc.). In order
to study the symmetries that arise from these structures we forgo this encoding.
More precisely, we consider computation models such as circuits whose inputs are themselves functions of type $K^X$ where we think of $X$ as a set of variables and $K$ a domain of values that the variables can take.  The set $X$ may have some further structure to reflect the intended structured input, but no more.  In particular, we do not assume that $X$ is linearly ordered.  For example, if $X =
V^2$ and $K = \{0, 1\}$ then the elements of $\{0, 1\}^X$ may be
naturally interpreted as directed graphs over the vertex set $V$.  Or, with the same $X$, if we let  $K = \ff$,  we can think of the elements of $K^X$ 
as matrices over $\ff$ with rows and columns indexed by $V$.

The symmetries of interest for a given class of structures correspond to group
actions on $X$, which lift to actions on $K^X$.  In this section we introduce the definitions of  $G$-symmetric functions, i.e.\ functions which are invariant under the action of
$G$ on its input, and $G$-symmetric
circuits, which are circuits computing $G$-symmetric functions but where the structure of the circuit itself and not just the output is invariant under the action of $G$.  The definitions are variations and generalizations of those from~\cite{DawarW20}.  We illustrate them with examples.

\subsection{Group Actions and Symmetric Functions}
\label{sec:group-actions-function}
\begin{definition}\label{def:symm-function}
  For a group $G$ acting on the domain of a function $F$ we say $F$ is
  $G$-\emph{symmetric} if for every $\sigma \in G$, $\sigma F = F$. We omit
  mention of the group when it is obvious from context.
\end{definition}

We are interested in functions of type $K^X \ra K$ with some group $G$ acting
on $X$, which then induces an action on $K^X$. We think of elements of $K^X$ as
``generalized structures'', and define notions of homomorphism and isomorphism
below. We first consider some examples. Note that whenever $H$ is a subgroup of
$G$, then any $G$-symmetric function is also $H$-symmetric.  In particular every function is $\{\id\}$-symmetric.

\begin{example}
  \label{eg:structs}
  \begin{enumerate}
  \item For a vertex set $V$, a function $F : \{0, 1\}^{V^2} \ra \{0, 1\}$
    defines a \emph{property} of (directed) graphs if it is $\sym_V$-symmetric,
    with the action of $\sym_V$ being defined simultaneously on both coordinates.
    We get properties of simple undirected graphs by considering
    $\sym_V$-symmetric functions $F : \{0, 1\}^{X} \ra \{0, 1\}$ where $X = {V
      \choose 2}$ is the collection of $2$-element subsets of $V$. Examples of
    graph properties include connectedness, Hamiltonicity and the existence of a
    perfect matching. A \emph{graph parameter} is a function $F : \{0, 1\}^{V^2}
    \ra \reals$ which is $\sym_V$-symmetric. Examples include the number of
    connected components, the number of Hamiltonian cycles and the number of
    perfect matchings.
  \item The elementary symmetric polynomial of degree $k$ in the set of
    variables $X$ is the polynomial:
    $$e_k(X) = \sum_{S \in {X \choose k}} \prod_{x \in S} x.$$
    For any field $\ff$, $e_k(X)$ defines a function $e_k^{\ff}: \ff^X \ra \ff$
    which is $\sym_X$-symmetric.
  \item If $X = \{x_{ij} \mid 1 \leq i,j \leq n\}$ is a \emph{matrix} of
    variables, then the trace $\tr(X)$, determinant $\det(X)$ and permanent
    $\perm(X)$ are polynomials that define, for any field $\ff$, functions of
    type $\ff^X$ which are $\sym_n$-symmetric where the group action is defined
    simultaneously on both coordinates.
  \item The permanent is invariant under separate row/column permutations. In
    other words, for $(\pi, \sigma) \in \sym_n \times \sym_n$, and matrix $M \in
    \ff^{n \times n}$, we have $\perm((M_{xy})) = \perm((M_{\pi(x)\sigma(y)}))$.
    So the permanent for $n\times n$-matrices is a $\sym_n \times
    \sym_n$-symmetric function.
  \item The trace and determinant are not $\sym_n \times \sym_n$-symmetric under
    the above action, but the symmetry group of the determinant is richer than
    just $\sym_n$. Define $D_n \leq \sym_n \times \sym_n$ to be the group $\{
    (\sigma,\pi) \mid \sgn{\sigma} = \sgn{\pi}\}$. This is a subgroup of $\sym_n
    \times \sym_n$ of index $2$. The determinant is $D_n$-symmetric and so, in
    particular $\alt_n \times \alt_n$-symmetric, since $\alt_n \times \alt_n
    \leq D_n$.
  \end{enumerate}
\end{example}

Let $G$ and $H$ be groups. A \emph{homomorphism} from the $G$-set $X$ to $H$-set
$Y$ is a pair of functions $(f, \phi)$ where $f: X \ra Y$ is a function and
$\phi: G \ra H$ is a group homomorphism such that for each $x \in X$ and $\pi
\in G$, $f(\pi x) = \phi(\pi) f(x)$. We say $(f, \phi)$ is an \emph{isomorphism}
if both $f$ and $\phi$ are bijective. We abuse terminology and refer to a
bijection $f : X \ra Y$ as an isomorphism if there exists $\phi : G \ra H$ such
that $(f, \phi)$ is an isomorphism.

Let $M : X \ra K$ with $X$ a $G$-set and let $N: Y \ra K$ with $Y$ an $H$-set. A
\emph{homomorphism} from $(M, G)$ to $(N, H)$ is a homomorphism $(f, \phi): (X,
G) \ra (Y, H)$ such that $N \circ f = M$. It is an \emph{isomorphism} from $(M,
G)$ to $(N,H)$ if both $f$ and $\phi$ are bijective. We omit mention of the
groups and refer just to a homomorphisms or isomorphisms from $M$ to $N$ when
the groups are clear from context (see Section~\ref{sec:bij-games}.

Let $G$ be a group acting on $X$ and $F: K^X \ra L$. We are usually interested
in such functions up to isomorphism. Let $M : Y \ra K$ and $H$ be a group acting
on $Y$ such that $(X, G)$ and $(Y, H)$ are isomorphic. Then we abuse notation
and sometimes write $F(M)$ to denote $F(M \circ f)$ for some isomorphism $f : X
\ra Y$.

\subsection{Symmetric Circuits}
\label{subsec:symmetric-circ}

We next define the notion of a symmetric circuit as it appears
in~\cite{DawarW20}. These circuits take as input functions of the form $M : X
\ra K$, where $X$ is a $G$-set, and are symmetric in the sense that the
computation itself, not just the value of the output, remains unchanged under
the action of $G$. We first need to formalize what it means for a permutation in
$G$ to act on the gates of a circuit, and for this we define the notion of a circuit
automorphism extending a permutation.

\begin{definition}[Circuit Automorphism]
  Let $C = (D, W)$ be a circuit over the basis $\BB$ with variables $X$ and
  values $K$. For $\sigma \in \sym_X$, we say that a bijection $\pi : D \ra D$
  is an \emph{automorphism} extending $\sigma$ if for every gate $g$ in $D$ we
  have that
  \begin{itemize}
  \item if $g$ is a constant gate then $\pi (g) = g$,
  \item if $g$ is a non-constant input gate then $\pi (g) = \sigma (g)$,
  \item if $(h,g) \in W$ is a wire, then so is $(\pi h, \pi g)$
  \item if $g$ is labelled by $b \in \BB$, then so is $\pi g$.
  \end{itemize}
\end{definition}

We say that a circuit $C$ with variables $X$ is \emph{rigid} if for every
permutation $\sigma \in \sym(X)$ there is at most one automorphism of $C$
extending $\sigma$. The argument used to prove~\cite[Lemma 5.5]{DawarW2020A}
suffices to show that any symmetric circuit may be transformed into an
equivalent rigid one in time polynomial in the size of the circuit. As such,
when proving lower bounds we often assume the circuit is rigid without a loss of
generality.

We are now ready to define the key notion of a symmetric circuit.

\begin{definition}[Symmetric Circuit]\label{def:symm-circuit}
  Let $G$ be a group acting on a set $X$ and $C$ be a circuit with variables
  $X$. We say $C$ is a \emph{$G$-symmetric circuit} if for every $\sigma \in G$
  the action of $\sigma$ on $X$ extends to an automorphism of $C$.
\end{definition}

The following can be shown via a straightforward induction.

\begin{prop}
  Let $C$ be a $G$-symmetric circuit. Then $C$ defines a $G$-symmetric function.
\end{prop}

%% file: games-and-supports.tex
Hella's bijection game~\cite{Hella1996} is a two-player game played on
relational structures, such as graphs. It was defined to demonstrate
indistinguishability of structures in logics with counting and extensions
thereof. The indistinguishability relations it defines on graphs are closely
tied to Weisfeiler-Leman equivalences~\cite{CFI92}. The games are played on a
pair of structures $A$ and $B$ by two players called Spoiler and Duplicator.
Spoiler aims to show that the two structures are different while Duplicator
pretends they are the same. We associate with each structure a sequence of $k$
pebbles which are placed in the course of the game on elements of the two
structures. The game proceeds in a sequence of rounds. The number of rounds can
be greater than $k$ and so pebbles can be moved from one element to another
during the course of the game. At each round, Spoiler chooses a pebble $p_i$ ($i
\in [k]$) from $A$ and the matching pebble $q_i$ from $B$. Duplicator provides a
bijection $h$ between the two structures. Spoiler then chooses an element $a$ of
$A$ on which to place $p_i$ and $q_i$ is placed on $h(a)$. In each round,
Duplicator must ensure that the partial map between the two structures defined
by the placement of the pebbles is a partial isomorphism, otherwise Spoiler
wins. For more details on this game see~\cite{grohe2017descriptive}.

The connection between bijection games and symmetric circuits is first made
in~\cite{AndersonD17} which showed a connection between the expressive power of
counting logics and symmetric Boolean circuits in the threshold basis $\BT$.
This leads to the suggestion (made in~\cite{Dawar2016}) of using bijection games
as a tool directly to prove circuit lower bounds. The main contribution of this
section is the generalization of these bijection games in two different
directions, as well as results establishing how this tool can be used
to prove more general circuit lower bounds.

We noted earlier that we may think of functions on $G$-sets as some sort of
generalised structure. The development of the theory of bijection games (and
supports) requires us to further restrict our attention to the case where $G$ is
a subgroup of some symmetric group. The domain of the symmetric group
corresponds in some sense to the universe of the structure in question. 

We also develop in Section~\ref{sec:supports} a theory of supports. The support
of a gate $g$ in a $G$-symmetric circuit for $G \leq \sym_A$ is a subset of $A$
which determines both the evaluation of $g$ and its orbit. We establish in
Section~\ref{sec:bounds-supports} the support theorem, which connects the
minimum size of these supports with the minimum size of the orbits of a gate
(and so the size of the circuit). We show in Section~\ref{sec:games-circuits}
that if Duplicator has a winning strategy in the game with $2k$ pebbles on a pair of indexed functions then these functions cannot be distinguished by any pair of $G$-symmetric circuits with supports of size at most $k$.  In Section~\ref{sec:bounds-supports} we combine this
result with the support theorem to show that to prove exponential lower bounds
it suffices to establish a linear lower bound on the number of pebbles needed
by Spoiler.

%% file: new-bijection.tex
We now introduce our generalization of Hella's bijection game.  The generalization is in two directions.  The first is that we allow
the game to be played on arbitrary indexed functions, rather than just graphs or
relational structures.  The second is that Duplicator is restricted on which bijections it is allowed to play.  Specifically, the bijection must be obtained from a fixed initial bijection by composition with a permutation from a fixed group.  We recover the usual requirement in the bijection game by just taking this group to be the full symmetric group.

We begin by introducing the key notions of indexed set and function.

\begin{definition}\label{def:indexed-set}
  Let $A$ be a set, $G \leq \sym_A$, and $X$ be a $G$-set. We call the triple
  $(X, A, G)$ an \emph{indexed set}. We call a triple $(M, A, G)$, where $M: X \ra K$ is
  a function on $X$, an \emph{indexed function}.
\end{definition}

All of the structures discussed so far may be identified with indexed functions.  For example, we can identify a (directed) graph $(V, E^2)$ with the indexed function $(E: V^2
\ra \{0, 1\}, V, \sym_V)$.  Similarly, a bipartitioned graph $(V, U, E)$ can be identified  with the indexed function $(E: U \times V \ra \{0, 1\}, U \uplus V, \sym_U \times \sym_V)$.

We now want to generalize the notion of isomorphism between structures to an equivalence notion on indexed functions, which is suitably restricted by the permutation group $G$.  Suppose $A$ and $B$ are index sets and $f: A \ra B$ is a bijection.  For a group $G \leq \sym_A$, let $H \leq \sym_B$ be the group of permutations $\{ f \pi f^{-1} \mid \pi \in G\}$.  Let $\mathcal{M} = (M:X \ra K, A, G)$ and $\mathcal{N} = (M:Y \ra K, B, H)$ be a pair of indexed functions and $\hat{f} : X \ra Y$ be a bijection which \emph{lifts} the bijection $f$ in the sense that $\hat{f}(\pi x) = (f \pi f^{-1})\hat{f}(x)$ for all $\pi \in G$.  We say that the indexed functions $\mathcal{M}$ and $\mathcal{N}$ are $(f,\hat{f})$-isomorophic, if there is a permutation $\pi \in H$ such that for all $x \in X$, $M(x) = N(\pi \hat{f}(x))$.

In particular, for a pair of directed graphs $(V_1,E_1)$ and $(V_2,E_2)$, if $f$ is any bijection between $V_1$ and $V_2$, and $\hat{f}$ is its natural lift to a bijection between $V_1^2$ and $V_2^2$ (i.e.\ $\hat{f}(u,v) = (f(u),f(v))$, then the pair of indexed functions $(E_1: V_1^2\ra \{0, 1\}, V_1, \sym_{V_1})$ and $(E_2: V_2^2\ra \{0, 1\}, V_2, \sym_{V_2})$ are $(f,\hat{f})$-isomorphic if, and only if,  $(V_1,E_1)$ and $(V_2,E_2)$ are isomorphic in the usual sense.  Similar statements hold for bipartitioned graphs and other relational structures.  As a further example, we can  identify $n\times n$ matrices over a field $\ff$ with
structured functions of the form $(M: [n] \times [n] \ra \ff, [n] \uplus [n],
G)$, for some group $G$.  Unlike the case of structures considered above, there is not a canonical choice of $G$, and it depends on what invariants of the matrix we wish to take into account.  For example, if $G = \sym_n$ then isomorphism corresponds to
equivalence under simultaneous row and column permutations and if $G = \sym_n
\times \sym_n$ then isomorphism corresponds to equivalence under separate
row and column permutations.

Our notion of isomorphism is parameterized by a fixed bijection between the index sets $A$ and $B$ and its lift to the indexed sets $X$ and $Y$.  To simplify notation, we often simply identify the sets $A$ and $B$, and the sets $X$ and $Y$, so that the base bijection can be taken to be the identity.  This is akin to fixing the universe of any $n$ vertex graph to be the set $[n]$, so that isomorphisms can be identified with permutations of $[n]$.

In order to define the bijection game, we need a notion of \emph{partial isomorphism} on the set of pebbled positions in an indexed function.  In order to introduce this, we first define the notion of a \emph{lift}.

\begin{definition}\label{def:lift}
  Let $(X, A, G)$ be an indexed set. For $S \subseteq A$ we define the
  \emph{lift} of $S$ to be the set $X_S = \{x \in X \mid \stab(S) \leq
  \stab(x) \}$.
\end{definition}

Thus, $X_S$ is those elements of $X$ that are fixed by any permutation that fixes $S$ pointwise.  The following is now an easy observation.

\begin{lemma}
  Let $(X, A, G)$ be an indexed set. If $X_S$ is the lift of $S \subseteq A$
  then $\setstab(S) \leq \setstab(X_S)$.
  \label{lem:lift-setstab}
\end{lemma}
\begin{proof}
  Let $\sigma \in \setstab(S)$, $\pi \in \stab(S)$, and $s \in S$. Then
  $\sigma^{-1}\pi\sigma (s) = s$, so $\sigma^{-1}\pi\sigma \in \stab(S)$.
  Hence, by definition of $X_S$, we have $\sigma^{-1}\pi\sigma(x) = x$ for any
  $x\in X_S$. Hence $\pi \sigma (x) = \sigma (x)$ and therefore $\pi \in
  \stab(\sigma(x))$. Since $\pi \in \stab(S)$ was arbitrary, $\sigma(x) \in
  X_S$ and therefore $\sigma \in \setstab(X_S)$.
\end{proof}

We define the notion of partial isomorphism specifically for two indexed functions over the same index set.  It can easily be generalized to the case with distinct index sets with a fixed bijection between them.

\begin{definition}\label{def:partial-iso}
  Let $(M: X \ra K, A, G)$ and $(N:X \ra K, A, G)$ be two indexed functions, and let $S \subseteq A$ and $\pi \in G$.  We say that $\pi$ induces a \emph{partial isomorphism} on $S$ if, for each $x \in X_S$, we have $M(x) = N(\pi x)$.
\end{definition}

It is easily seen that this gives the usual notion of partial isomorphism when the indexed functions are graphs or other relational structures.  With this, we are ready to define the bijection game on indexed functions.  The game is played by two players Spoiler and Duplicator on a pair of indexed functions, using a set of \emph{pebbles}.  During the course of the game, the pebbles are placed on elements of the indexing sets.  Where it does not cause confusion, we do not distinguish notationally between the pebbles and the elements on which they are placed.

\begin{definition}
  Let $\mathcal{N} = (M: X \ra K, A, G)$ and $\mathcal{N} = (N: X \ra K, A, G)$ be indexed functions.  The \emph{$k$-pebble $G$-bijection game} on $(M,N)$ is defined as follows.   The game
  is played between two players called the \emph{Spoiler} and \emph{Duplicator}
  using two sequences of pebbles $a_1, \ldots, a_k$ and $b_1, \ldots, b_k$. In each round of the game:
  \begin{enumerate}
  \item Spoiler picks a pair of pebbles $a_i$ and $b_i$,
  \item Duplicator picks a permutation $\pi \in G$ such that for each $j \neq
    i$, $\pi(a_j) = b_j$, and
  \item Spoiler chooses some $a \in A$ and places $a_i$ on $a$ and $b_i$ on
    $\pi(a)$.
  \end{enumerate}
  Spoiler has won the game  at the end of the round if the permutation $\pi$ 
does not induce a partial  isomorphism on the set $\{a_1,\ldots,a_k\}$.  We say that Duplicator has a winning strategy for the $k$-pebble $G$-bijection game
if it has a strategy to play the game forever without Spoiler winning.
\end{definition}

In the sequel, we abbreviate ``$k$-pebble $G$-bijection game'' to \emph{$(G,k)$-bijection game}.

We recover the ordinary bijection game on graphs by taking $A = [n]$, $X = [n]^2$ and $G = \sym_n$.  It is clear in this case that if $\mathcal{M}$ and $\mathcal{N}$ are isomorphic graphs, then Duplicator has a winning strategy by choosing a fixed isomorphism $\pi$ between them and playing it at every move.  However, if $G \leq \sym_A$ is a more restricted group that does not contain an isomorphism between $\mathcal{M}$ and $\mathcal{N}$, a Duplicator winning strategy is not guaranteed even when $\mathcal{M}$ and $\mathcal{N}$ are isomorphic as graphs.  This is just the case in the application in
Section~\ref{sec:lower-bound}.

%% file: supports.tex
Lower bounds for symmetric circuits rely on the notion of a \emph{support}. We
define the notion formally below but it is worthwhile developing an intuition.
If we have a $G$-symmetric circuit $C$ then the function computed by $C$ is
invariant under permutations in $G$ on the input.  This is not the case for individual gates
$g$ in $C$ other than the output gate: applying a permutation $\pi \in
G$ to the inputs of the circuit might
yield a different function at $g$.  But, by the symmetry condition, there is then
another gate $\pi g$ which computes this other function.  If the circuit is
small, the orbit of $g$ is small and thus the stabilizer group of $g$ is large.
That is, for many $\pi$ we have $g = \pi g$.  What the support theorem tells us
is that in this case, there is a small subset $S$ (which we call a
\emph{support} of $g$) of the permutation domain of $G$
such that the function computed at $g$ only depends on $S$.  This is in the sense
that permutations that only move elements of the permutation domain outside of
$S$ do not change the function computed at $g$.  This support theorem can be
proved as long as the group $G$ is, in a sense, large enough.  We now define the
notion of a support formally and prove the relationship between support size and
orbit size in Section~\ref{sec:games-circuits}.

\begin{definition}
  Let $X$ be a $G$-set and $H \leq G$ a subgroup of $G$.  We say that
  $S \subseteq X$ is a \emph{support} of $H$   if $\stab_G (S) \leq H$.
\end{definition}

We extend this notion to indexed sets, where the support is now a
subset of the indices.  A key example is the action of $\sym_n$ on the set
$X = [n]\times[n]$ or on the collection of matrices $M: X \ra \{0,1\}$.

\begin{definition}
  Let $(X, A, G)$ be an indexed set. We say that a set $S \subseteq A$ is a
  support of $x \in X$ if it is a support of $\stab(x)$.
\end{definition}

Note that $S$ is a support of $x$ just in case $x$ is in the lift of
$S$, in the sense of Definition~\ref{def:lift}.

Let $(X, A, G)$ be an indexed set.  We write $\ORB(X)$ to denote the
maximum size of the orbit of $x$ over all $x \in X$.  For $x \in X$ we
write $\MSP(x)$ to denote the minimum size of a support of $x$ and
$\SP(X)$ to denote the maximum of $\MSP(x)$ over all $x \in
X$.

We now specialise our discussion to circuits. Let $C$ be a rigid
$G$-symmetric circuit with gate set $D$, where $G \leq \sym_n$ for some natural number $n$. Then
$(D, [n], G)$ is an indexed set.  In this way we can speak of the supports and
orbits of a gate.  We abuse notation slightly and write $\ORB(C)$ and $\SP(C)$ to
denote $\ORB(D)$ and $\SP(D)$, respectively.

%% file: games-on-circuits.tex
We are now ready to prove the first major theorem of this section. This links
the number of pebbles in a bijection game with the size of the
supports.
Structures in which Duplicator has a $2k$-pebble winning strategy cannot be
distinguished by symmetric circuits with supports limited to size $k$. The
statement of the theorem and argument used generalize that in~\cite{Dawar2016}.

\begin{theorem}
  Let $(X, A, G)$ be an indexed set and let $C$ be a $G$-symmetric circuit with
  values from $K$ and variables $X$, such that $\SP(C) \leq k$.   If Duplicator has a winning strategy
  for the $(G, 2k)$-bijection game played on $(M,A,G)$ and $(N,A,G)$
  for functions $M : X \ra K$ and $N: X  \ra K$ then $C[M] = C[N]$.
  \label{thm:game-circuits}
\end{theorem}

\begin{proof}
  We prove this result by contraposition, i.e.\ we assume $C[M] \neq C[N]$ and
  show that Spoiler has a winning strategy for the $(G, 2k)$-game on
  $(M, N)$.

  The strategy we define for Spoiler can be understood informally as follows. We
  start at the output gate $g$ and note that for any bijection $\sigma \in G$, $C[M](g)
  \neq C[\sigma N](g)$. The goal of Spoiler now over the next (at most) $k$
  rounds is to pebble a support of some child $h$ of $g$ such that for
  Duplicator's most recent chosen bijection $\sigma$ we have that $C[M](h) \neq
  C[\sigma N](h)$. Since the bijection chosen by Duplicator can be interpreted
  as a permutation $\sigma \in G$ respecting the pebble configuration, and the
  fact that an entire support of $h$ is pebbled and so any such permutation
  fixes $h$, it follows that $C[M](h) \neq C[\sigma N](h)$ for any choice of
  $\sigma$. We now iterate this process, aiming next to pebble a support of an
  appropriate child of $h$ over the next (at most) $k$ rounds. We terminate when
  we reach an input gate, which witnesses that Spoiler has won the
  game.

  So, fix for each gate $g$ of $C$ a support of size at most $k$.  We
  write $\consp(g)$ to denote this support.  

  \begin{claim}\label{clm:strategy}
    Suppose the pebbled position is $\vec{a}$ and $\vec{b}$. Let $g$ be a gate
    such that $\consp(g) \subseteq \vec{a}$ and for some $\sigma$ that maps
    $\vec{a}$ to $\vec{b}$ we have that $C[M](g) \neq C[\sigma N](g)$. There
    exists a strategy for Spoiler such that after at most $k$ rounds where the
    pebbled positions are now $\vec{a}'$ and $\vec{b}'$ there is 
    $h \in \child{g}$ such that $\consp(h) \subseteq \vec{a}'$ and for all $\sigma' \in G$ that maps
    $\vec{a}'$ to $\vec{b}'$ we have that $C[M](h) \neq C[\sigma' N](h)$.
  \end{claim}

  \begin{proof}
    We describe an invariant that Spoiler is able to maintain inductively.
    Suppose that after $i \geq 0$ rounds $c_1, \ldots, c_i$ and $d_1, \ldots,
    d_i$ have been pebbled. Let $S_i$ be the pointwise stabiliser of $c_1,
    \ldots, c_i$ in $\stab(\vec{a})$. Then $S_0 = \stab(\vec{a})$. Spoiler ensures there is a
    gate $h_i \in \child{g}$ with $c_1, \ldots, c_i \in \consp(h_i)$ and a $\alpha \in
    K$ such that

    \begin{align*}
      \vert \{ h \in \orb_{S_i}(h_i) \mid C[M](h) = \alpha\} \vert \neq  \vert \{h \in \orb_{S_i}(h_i) \mid C[\sigma N ](h) = \alpha \} \vert
    \end{align*}
    for all $\sigma \in G$ that maps $\vec{a}, c_1, \ldots, c_i$ to $\vec{b},
    d_1, \ldots, d_i$. This suffices to prove the claim since $\vert
    \consp(h_i) \vert \leq k$ and therefore for some $i$, we will have
    $i \geq \vert    \consp(h_i) \vert$ and in this case,
    $\orb_{S_i}(h_i)$ is a singleton.

    We first consider the case $i = 0$. We have for any $\sigma \in G$ that maps
    $\vec{a}$ to $\vec{b}$ that $C[M](g) \neq C[\sigma N](g)$.  Since
    the  basis element labelling $g$ is a fully symmetric function,
    there must be some $\alpha \in K$ such that $\vert\{ h \in \child{g} \mid C[M]h
  = \alpha \}\vert$ is different from $\vert\{ h \in \child{g} \mid C[\sigma N]h
  = \alpha \}\vert$.  Moreover, since $S_0 \leq \stab(g) \leq
  \setstab(\child{g})$, $S_0$  partitions $\child{g}$ into orbits.  It
    follows that for some $h \in \child{g}$
    \begin{align*}
      \vert \{ h' \in \orb_{S_0}(h) \mid C[M](h') = \alpha\} \vert \neq  \vert \{h' \in \orb_{S_0}(h) \mid C[\sigma N](h') = \alpha \} \vert.
    \end{align*}
    Now, take any $\sigma' \in G$
    that maps $\vec{a}$ to $\vec{b}$. Let $\rho := \sigma^{-1}\sigma' $. Then
    $\rho \in S_0$, and so $\orb_{S_0}(h) = \orb_{S_0}(\rho h)$.  Moreover, for any $h' \in
    \orb_{S_0}(h)$, $C[\sigma N ] (h') = C[\sigma\rho N] (\rho h') = C[\sigma'
    N](\rho h')$, and so
    \begin{align*}
      \vert \{ h' \in \orb_{S_0}(h) \mid C[M](h') = \alpha\} \vert \neq  \vert \{h' \in \orb_{S_0}(h) \mid C[\sigma'N ](h') = \alpha \} \vert.
    \end{align*}
    Thus, taking $h_0 := h$ suffices.

    Inductively, suppose Spoiler has maintained the invariant after $i > 0$
    rounds. At the start of round $i+1$ Spoiler picks up a pair of pebbles
    $(a_j, b_j)$ where $a_j$ is not placed on $c_1, \ldots, c_i$ or an element
    of $\consp(g)$. Suppose Duplicator plays a bijection $\sigma_{i+1}$. Then,
    by the induction hypothesis there is an $h_i$ and an $\alpha_i$ such that
    \begin{align*}
      \vert \{ h' \in \orb_{S_i}(h_i) \mid C[M](h') = \alpha_i\} \vert
      \neq  \vert \{h' \in \orb_{S_i}(h_i) \mid C[\sigma_i N  ](h') = \alpha_i \} \vert.
    \end{align*}
    and $c_1, \ldots, c_i \in \consp(h_i)$. Let $c_1, \ldots, c_i, s_{i+1},
    \ldots, s_u$ enumerate $\consp(h_i)$. Then $\orb_{S_i}(h_i)$ is partitioned
    into sets $(O_c)_{c\in A \setminus \vec{a} \cup \{c_1, \ldots, c_i\}}$,
    where $O_c = \{\pi h_i \mid \pi (s_{i+1}) = c\}$. Thus there is a value
    $c_{i+1}$ and $\alpha_{i+1}$ for which
    \begin{align*}
      \vert \{ h' \in O_{c_{i+1}} \mid C[M](h') = \alpha_{i+1}\} \vert \neq  \vert \{h' \in O_{c_{i+1}} \mid C[\sigma N](h') = \alpha_{i+1} \} \vert.
    \end{align*}

    Spoiler then places $a_j$ on $c_{i+1}$ and $b_j$ on $\sigma_{i+1}(c_{i+1})$.
    Let $h_{i+1}$ be any element of $O_{c+1}$. The required invariant is
    satisfied by construction.
  \end{proof}
  
  It follows from Claim~\ref{clm:strategy} that in at most $k \cdot \depth
  (C)$ rounds Spoiler can force the pebbles to a position such that for some
  input gate $g$ labelled by some variable $x \in X$ we have: that for any
  $\sigma \in G$ that Duplicator may play we have that $C[M ](g) \neq C[\sigma
  N ](g)$. Thus $M (x) \neq \sigma N (x)$, and so Spoiler wins.

\end{proof}

%% file: bounds-on-supports.tex
The main result of this subsection establishes that, for suitable
choice of groups $G$, if a family of $G$-symmetric
circuits has subexponential size orbits than it has sublinear size supports. We
prove this specifically for the case when $G$ is an alternating group. The
argument extends easily to cases where $G$ contains a large alternating group,
and we derive one such instance in
Corollary~\ref{cor:support-theorem-cor}.  The
proof relies on a standard fact about permutation groups, attributed to Jordan
and Liebeck, that translates a bound on orbit size to a bound on support size.
To understand this, suppose $g$ is a gate with orbit size (relative to $\alt_n$)
bounded by ${n \choose k}$. Then by the orbit-stabilizer theorem, this is
equivalent to $[\alt_n : \stab(g)] \leq {n \choose k}$. One way for $\stab(g)$
to have small index in this way is for it to always contain a large copy of the
alternating group. That is $\stab(g)$ contains the alternating group restricted
to $n-k$ elements, or equivalently $g$ has a support of size at most $k$.
Theorem~\ref{thm:dixonmort} asserts that indeed this is the only way.

\begin{theorem}[\cite{dixon1996permutation}, Theorem 5.2B]
  Let $Y$ be a set such that $n := \vert Y \vert > 8$, and let $k$ be an integer
  with $1 \leq k \leq \frac{n}{4}$. Suppose that $G \leq \alt_Y$ has index
  $[\alt_Y:G] < {n \choose k}$ then there exists $X \subset Y $ with $\vert X
  \vert < k$ such that $\stab_{\alt_Y}(X) \leq G$.
  \label{thm:dixonmort}
\end{theorem}

We derive from Theorem~\ref{thm:dixonmort} the following asymptotic relationship
between orbit and support size.  An analogous version of this with respect to the
symmetric group is stated in the full version of~\cite{DawarW20} and
the proof is largely the
same. We include a proof for completeness.

\begin{theorem}
  Let $(C_n)$ be a family of rigid $\alt_n$-symmetric circuits. If $\ORB(C_n) =
  2^{o(n)}$ then $\SP(C_n) = o(n)$.
  \label{thm:support-theorem}
\end{theorem}
\begin{proof}
  Let $k$ be the least value such that $\ORB{C_n}  \leq {n \choose
    k}$.  By the assumption that $\ORB(C_n) = 2^{o(n)}$, we have that
  $k$ is $o(n)$.  Indeed, otherwise there is a constant $c$ with $0 <
  c < \frac{1}{2}$, such that $k -1 \geq cn$ for infinitely many $n$.
  And since ${n \choose l} \geq (\frac{n}{l})^{l}$ for all $l$ it
  follows that ${n \choose {k-1} } \geq (\frac{n}{cn})^{cn} > 2^{cn}$.
  Since $k$ is the \emph{least} value such that $\ORB(C_n)  \leq {n
    \choose k}$, it follows that $\ORB(C_n) \geq 2^{cn}$ for
  infinitely many $n$, contradicting the assumption that $\ORB(C_n) =
  2^{o(n)}$.

  From $k = o(n)$ it follows that for all large enough $n$, $k \leq
  \frac{n}{4}$.  Then, for any gate $g$ of $C_n$, by the
  orbit-stabilizer theorem, we have $[\alt_n:\stab(g)] \leq {n \choose
    k}$ and so by Theorem~\ref{thm:dixonmort}, $g$ has a support of
  size less than $k$.
\end{proof}

The following corollary establishes the analogue of
Theorem~\ref{thm:support-theorem} needed to prove our lower bounds for $\alt_n
\times \alt_n$-symmetric circuits.

\begin{cor}
  Let $(C_n)_{n \in \nats}$ be such that for each $n$, $C_n$ is defined over the
  matrix of variables $X_n = \{x_{i, j} \mid i, j \in [n]\}$ and $C_n$ is a
  rigid $\alt_n \times \alt_n$-symmetric circuit. If $\ORB(C_n) = 2^{o(n)}$ then
  $\SP(C_n) = o(n)$.
  \label{cor:support-theorem-cor}
\end{cor}
\begin{proof}
  Fix $G = \alt_n \times \alt_n$ and let $G_r = \alt_n \times \{\id\} \leq G$
  and $G_c= \{\id\} \times \alt_n \leq G$ be the restriction of the action of
  $G$ to the two coordinates. We can think of them as the actions of $G$ on the
  rows and columns respectively.  For any gate $g$ in $C_n$, let $S_r$ and $S_c$
  be supports of $g$ under the action of $G_r$ and $G_c$ respectively. Then
  it is easily seen that $S_r \cup S_c$ is a support of $g$.

  Suppose for all gates $g$ in $C_n$ we have $\ORB(C_n) = 2^{o(n)}$. Since the
  orbit of $g$ under the action of $G$ includes its orbits under the action of
  its subgroups $G_r$ and $G_c$, we have that each of $\orb_{G_r}(g)$
  and $\orb_{G_c}(g)$ has size $2^{o(n)}$ and therefore by
  Theorem~\ref{thm:support-theorem} $S_r$ and
  $S_c$ can be chosen of  size $o(n)$ and the result follows.
\end{proof}

We now combine these results along with Theorem~\ref{thm:game-circuits} to
establish the crucial connection between bijection games and exponential lower
bounds for symmetric circuits. We prove the result for the particular case of
interest to us in this paper, namely for $\alt_n \times \alt_n$-symmetric
circuits taking as input $n\times n$-matrices, but note that it holds more
generally.

\begin{theorem}
  Fix a set $K$, and for each $n \in \nats$, let $G_n$ denote the
  group $\alt_n \times  \alt_n$.  Fix  a $G_n$-set $X_n$ and   let $P = (P_n)_{n \in \nats}$ be a family of
  functions $P_n : K^{X_n} \ra K$, where $P_n$ is $G_n$-symmetric.  Suppose there are infinitely many $n$ for which  there are pairs $M_{n}, N_{n} : X_{n} \ra K$ such that $P_n(M_n)
  \neq P_n(N_n)$ and Duplicator has a strategy to win the
  $(G_n, k)$-bijection game played on $(M_n,[n]\uplus
  [n], \alt_n\times \alt_n)$ and $(N_n,[n]\uplus [n], \alt_n\times
  \alt_n)$ for $k = \Omega(n)$.  Then there is no family of
  $G_n$-symmetric circuits that  computes $P$ and has size
  $2^{o(n)}$. 
  \label{thm:general-lower-bound}
\end{theorem}
\begin{proof}
  From Theorem~\ref{thm:game-circuits} any $G_n$-symmetric circuit $C_n$ that
  has $\SP(C_n) \leq k/2$ must have that $C[M_{n}] = C[N_{n}]$, and so
  cannot compute $P_n$.  It follows then that any family of $G_n$-symmetric
  circuits that computes $P_n$ must have $\SP(C_n) = \Omega(n)$ and so from
  Corollary~\ref{cor:support-theorem-cor} cannot have orbits of size $2^{o(n)}$,
  and hence cannot have size $2^{o(n)}$.
\end{proof}

Notice that both the definition of the bijection games and
Theorem~\ref{thm:game-circuits} place almost no restriction on the group actions
considered. The link between the number of pebbles in the game and the size of
the support is robust. However, the application in
Theorem~\ref{thm:general-lower-bound} is for a severely limited group action.
This is because the link between support size and orbit size proved in
Theorem~\ref{thm:support-theorem} requires the presence of a large alternating
group.

%% file: lower-bound.tex
We now deploy the machinery we have developed to prove the main lower
bound result of this paper.

\begin{theorem}[Main Theorem]
  Let $\ff$ be a field of characteristic $0$. There is no family of $\alt_n
  \times \alt_n$-symmetric circuits $(C_n)_{n \in \nats}$ over $\ff$ of size
  $2^{o(n)}$ computing the determinant over $\ff$.
  \label{thm:main}
\end{theorem}

To prove 
Theorem~\ref{thm:main} we need to
construct for each $k$, a pair of $n \times n$ $\{0,1\}$-matrices $M_k$ and
$N_k$ with $n = O(k)$, $\det(M_k) \neq \det(N_k)$ and such that Duplicator has a
winning strategy in the $(\alt_n \times \alt_n,k)$-bijection game played
on $M_k$ and $N_k$.

We construct the matrix $M_k$ as the biadjacency matrix of a bipartite graph
$\Gamma$. The matrix $N_k$ is obtained from $M_k$ by interchanging a pair of
columns of $M_k$. Hence, $N_k$ is also a biadjacency matrix of $\Gamma$ and
$\det(M_k) = - \det(N_k)$ by construction. Thus, as long as $\det(M_k) \neq 0$
the two determinants are different.

In Section~\ref{sec:graph} we describe the construction of the graph which gives
rise to the biadjacency matrix $M_k$. This graph is obtained by a CFI
construction from a base graph $\Gamma$ satisfying a number of conditions. We
show the existence of graphs $\Gamma$ satisfying these conditions in
Section~\ref{sec:base-graph}. Then, in Section~\ref{sec:game} we argue that
Duplicator has a winning strategy in the $(\alt_n \times \alt_n,k)$-bijection game played on $M_k$ and $N_k$.  
We bring it all
together in Section~\ref{sec:main-proof} for a proof of Theorem~\ref{thm:main}.

\subsection{Constructing the Graph}\label{sec:graph}
In proving the lower bound for symmetric circuits computing the permanent
in~\cite{DawarW20}, we adapted a construction due to Cai, F\"{u}rer and
Immerman~\cite{CFI92} of pairs of graphs not distinguished by the
$k$-dimensional Weisfeiler-Leman algorithm. We showed that we could obtain such
a pair of bipartite graphs with different numbers of perfect matchings. Note
that saying a pair of graphs are not distinguished by the $k$-dimensional
Weisfeiler-Leman algorithm is the same as saying that Duplicator has a winning
strategy in the $(k+1)$-pebble bijection game, using the full
symmetric group. In
the present construction, we consider a game played on a pair of isomorphic
bipartite graphs but where the set of permissible bijections does not include
any isomorphisms between them. Equivalently, we play the game on two distinct
biadjacency matrices for the same graph. The graphs we consider are, indeed,
exactly the graphs used in~\cite{CFI92} except that we have to ensure they are
bipartite.

\begin{figure}[h]
  \centering{ 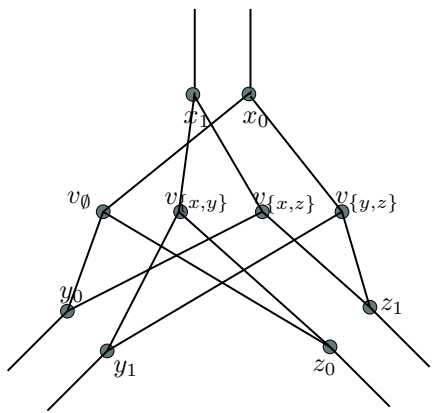
    \caption{A gadget in $\hG$ corresponding to vertex $v$ with neighbours
      $x,y,z$}
    \label{fig:gadget}
  }
\end{figure}

\paragraph*{CFI graphs and Determinants}
Let $\Gamma = (U \cup V, E)$ be a $3$-regular bipartite graph with bipartition
$U,V$. We obtain the graph $\hG$ by replacing each vertex $v$, with neigbours
$x,y,z$ with the ten-vertex gadget depicted in Figure~\ref{fig:gadget}. This
gadget is described as follows.
\begin{itemize}
\item There is a set denoted $I_v$ of four \emph{inner vertices}: a vertex $v_S$
  for each set $S \subseteq \{x,y,z\}$ of even size.
\item There is a set denoted $O_v$ of six \emph{outer vertices}: two $u_0,u_1$
  for each $u \in \{x,y,z\}$.
\item There is an edge between $v_S$ and $u_1$ if $u \in S$ and an edge between
  $v_S$ and $u_0$ if $u \not\in S$.
\end{itemize}
Corresponding to each edge $e = \{u,v\} \in E$ there is a pair of edges that we
denote $e_0$ and $e_1$ in $\hG$: $e_0$ connects the vertex $u_0 \in O_v$ with
$v_0 \in O_u$ and $e_1$ connects the vertex $u_1 \in O_v$ with $v_1 \in O_u$

Note that the graph $\hG$ is bipartite. Indeed, if we let $X := \bigcup_{v \in
  U}I_v \cup \bigcup_{v \in V} O_v$ and we let $Y := \bigcup_{v \in V}I_v \cup
\bigcup_{v \in U} O_v$ then it is easily seen that all edges in $\hG$ are
between $X$ and $Y$. Since $\Gamma$ is a $3$-regular bipartite graph, it follows
that $|U| = |V|$ and therefore $|X| = |Y|$. Writing $m$ for $|U|$ and $n$ for
$|X| = 10m$, we obtain a biadjacency $n \times n$ matrix representing the graph
$\hG$ by fixing a pair of bijections $\eta: X \ra [n]$ and $\eta': Y \ra [n]$.
The action of the group $D_n$ divides the collection of all such matrices into
two orbits. Letting $M$ and $N$ be representatives of the two orbits, we have
$\det(M) = -\det(N)$. We next aim to show that $\det(M) \neq 0$, provided that
$\Gamma$ has an odd number of perfect matchings.

Suppose we are given bijections $\eta: X \ra [n]$ and $\eta': Y \ra
[n]$ which together
determine a biadjacency matrix $M$ representation of $\hG$. We can also identify
each perfect matching in $\hG$ with a bijection $\mu: X \ra Y$ such that
$\mu(x)$ is a neighbour of $x$ for all $x \in X$. We write $\mtch(\hG)$ for the
collection of all perfect matchings of $\hG$. Then, the determinant of $M$ is
given by:
\begin{eqnarray*}
  \det(M) & = & |\{ \mu \mid \mu \in \mtch(\hG) \text{ with } \sgn{\eta'\mu\eta^{-1}} = 1 \}| \\ & & - |\{ \mu \mid \mu \in \mtch(\hG) \text{ with } \sgn{\eta'\mu\eta^{-1}} = -1 \}|.
\end{eqnarray*}
From now on, we take $\eta$ and $\eta'$ to be fixed and write
$\sgn{\mu}$ and talk of the sign of a matching $\mu$ as short hand for $\sgn{\eta'\mu\eta^{-1}}$.

To show that $\det(M)$ is non-zero, we analyze the structure of
the set $\mtch(\hG)$.  Note that since $\eta$ and $\eta'$ are
fixed, this imposes a linear order on the sets $X$ and $Y$. It also
induces a linear order on the sets $U$ and $V$, for instance by saying that $u <
v$, for $u,v \in U$ if the least element of $\eta(I_u \cup O_u)$ is less than
the least element of $\eta(I_v \cup O_v)$. We make use of this order without
further elaboration.

  \paragraph*{Perfect Matchings}
  In any perfect matching $\mu$ of $\hG$, all four vertices in $I_v$ for any $v
  \in U \cup V$ must be matched to vertices in $O_v$ as they have no neighbours
  outside this set. Thus, exactly two of the vertices of $O_v$ are matched to
  vertices in other gadgets. These two could be two vertices in a pair, e.g.\
  $\{x_0,x_1\}$ in Figure~\ref{fig:gadget} or they could be from different pairs, e.g.\ $\{x_0,y_0\}$. It
  is easily checked, by inspection of the gadget, that in either case, removing
  the two vertices of $O_v$ from the gadget results in an 8-vertex graph that
  admits a perfect matching. Indeed, if we remove two vertices in a pair, such
  as $\{x_0,x_1\}$ the resulting graph is a two-regular bipartite graph on eight
  vertices. The graph resulting from removing $x_0$ and $y_0$ from the gadget is
  depicted in Figure~\ref{fig:gadget-matching} and all other cases of removing
  two vertices of $O_v$ from different pairs result in a graph isomorphic to
  this one. We now classify perfect matchings in $\mtch(\hG)$ according to which
  edges \emph{between} gadgets are included in the matching.

  Say that a matching $\mu \in \mtch(\hG)$ is \emph{uniform} if for each edge $e
  \in E$, at most one of the two edges $e_0$ and $e_1$ is included in $\mu$.
  Thus, $\mu$ is \emph{non-uniform} if for some $e \in E$ both $e_0$ and $e_1$
  are included in $\mu$. Our first aim is to show that the non-uniform matchings
  contribute a net zero to the determinant of $M$.  Write
  $\umtch(\hG)$ to denote the set of uniform matchings of $\hG$.
  \begin{lemma}\label{lem:non-uniform}
    \begin{eqnarray*}
      \det(M) & = & |\{ \mu \mid \mu \in \umtch(\hG) \text{ with } \sgn{\mu} = 1 \}| \\ & & - |\{ \mu \mid \mu \in \umtch(\hG) \text{ with } \sgn{\mu} = -1 \}|.
    \end{eqnarray*}
  \end{lemma}
  \begin{proof}
    For a non-uniform matching $\mu$, let $e \in E$ be an edge for which both
    $e_0$ and $e_1$ are included in $\mu$ and and $v \in U \cup V$ a vertex such
    that $e$ is incident on $v$. Then, the four vertices in $I_v$ are matched
    with the remaining four vertices in $O_v$ and there are exactly two ways
    this can be done. To see this, consider the gadget in
    Figure~\ref{fig:gadget} and suppose that $e = \{v,x\}$ so $x_0$ and $x_1$
    are matched outside the gadget. Then the only two possible matchings of the
    remaining eight vertices are $v_{\emptyset} - y_0; v_{\{x,y\}}-z_0;
    v_{\{y,z\}}-y_1; v_{\{x,z\}}-z_1$ and $v_{\emptyset} - z_0; v_{\{x,y\}}-y_1;
    v_{\{y,z\}}-z_1; v_{\{x,z\}}-y_0$. Note that the second one is obtained from
    the first by composing with an \emph{odd} permutation: namely the $4$-cycle
    $(y_0 z_0 y_1 z_1)$. We can then define an involution on the set of
    non-uniform matchings as follows: map $\mu$ to the unique non-uniform
    matching $\mu'$ which differs from $\mu$ only in the set $O_v$ corresponding
    to the vertex $v \in U$ which is minimal (in the order on $U$) among all
    vertices of $U$ incident on some edge $e$ for which $e_0,e_1$ are in $\mu$.
    This is easily seen to be an involution by our previous observation.
    Moroever, it takes any matching $\mu$ to one of opposite sign.

    We conclude that all non-uniform matchings $\mu$ come in pairs of opposite
    sign and therefore cancel out in the expression for the determinant of $M$.
  \end{proof}

    \paragraph*{Uniform Perfect Matchings.}
  Our next aim is to count uniform perfect matchings in $\hG$ and classify them
  by sign. Suppose then that $\mu$ is a perfect matching that includes for each
  $e \in E$ at most one of the two edges $e_0$ and $e_1$. Let $F_\mu \subseteq
  E$ be the set of those $e \in E$ such that \emph{exactly} one of $e_0$ and
  $e_1$ is in $\mu$. Furthermore, let $f_\mu: F_\mu\ra\{0,1\}$ be the function
  given by $f_{\mu}(e) = i$ if, and only if, $e_i$ is in $\mu$.

  Since for each $v \in U \cup V$, exactly two of the vertices of $O_v$ are
  matched to vertices in other gadgets we can see that $F_\mu$ includes exactly
  two edges incident on every vertex $v$. In other words, $F_\mu$ is a
  $2$-factor of $\Gamma$ and therefore has exactly $2m$ edges.

   For a $2$-factor $F$ of $\Gamma$ and a function $f: F \ra \{0,1\}$, write
  $\mu(F,f)$ for the collection of all matchings $\mu$ of $\hG$ with $F_{\mu} =
  F$ and $f_{\mu} = f$.
  \begin{lemma}\label{lem:fixed-Ff}
    There are exactly $2^{2m}$ perfect matchings in $\mu(F,f)$, for any
    $2$-factor $F$ of $\Gamma$ and any function $f: F \ra \{0,1\}$.
  \end{lemma}
  \begin{proof}
    Let $v$ be a vertex of $\Gamma$ with neighbours $x,y,z$ and consider the
    gadget in Figure~\ref{fig:gadget}. Exactly two of the edges incident on $v$
    are included in $F$ and suppose that it is the edges $\{v,x\}$ and
    $\{v,y\}$. Further, suppose $f(\{v,x\}) = f(\{v,y\}) =0$. This means that
    the matching must pair the vertices $x_0$ and $y_0$ with vertices outside
    the gadget and we verify that the gadget with these two vertices removed
    admits two distinct perfect matchings. The gadget with vertices $x_0$ and
    $y_0$ removed is pictured in Figure~\ref{fig:gadget-matching}, where for
    clarity, we have removed the set brackets in the subscripts of the vertex
    labels. It is clear that $v_{\emptyset}$ must be matched with $z_0$. This
    leaves a six-cycle $v_{\{x,y\}}-x_1-v_{\{x,z\}}-z_1-v_{\{y,z\}}-y_1$ which
    admits two matchings. Entirely analogously, if $f(\{v,x\}) = f(\{v,y\}) =1$
    we can consider the gadget with the vertices $x_1$ and $y_1$ removed and it
    is easily checked that the resulting graph is isomorphic to the one depicted
    in Figure~\ref{fig:gadget-matching}, as is also the case when $f(\{v,x\})
    \neq f(\{v,y\})$.
    \begin{figure}[h]
      \centering{ 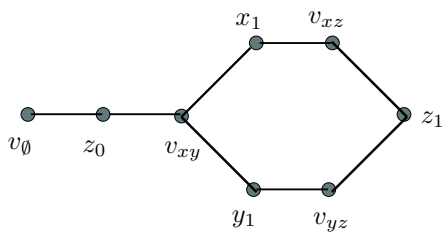
        \caption{The gadget from Figure~\ref{fig:gadget} with $x_0$ and $y_0$
          removed}
        \label{fig:gadget-matching}
      }
    \end{figure}
    Since the choice of the matching at each gadget is independent, and there
    are $2m$ gadgets, one for each vertex in $U \cup V$, we see that there are
    $2^{2m}$ distinct matchings $\mu$ for the fixed choice of $2$-factor $F$ and
    function $f: F \ra \{0,1\}$.
  \end{proof}

  The proof of Lemma~\ref{lem:fixed-Ff} shows that the two matchings $\mu$ and
  $\mu'$ in $\mu(F,f)$ obtained by varying the choice at just one gadget are
  related to each other by an \emph{even} permutation. For instance, the two
  matchings in Figure~\ref{fig:gadget-matching} are related by the $3$-cycle
  $x_1y_1z_1$. We can immediately conclude that all $2^{2m}$ matchings in
  $\mu(F,f)$ have the same sign.
  \begin{lemma}\label{lem:sign}
    For any $2$-factor $F$ of $\Gamma$, any function $f: F \ra \{0,1\}$ and any
    matchings $\mu_1,\mu_2 \in \mu(F,f)$, $\sgn{\mu_1} =
    \sgn{\mu_2}$.
  \end{lemma}

  We next show that we can go further and show that the sign of any matching in
  $\mu(F,f)$ does not depend on the choice of $f$.
  \begin{lemma}\label{lem:sign-factor}
    For any $2$-factor $F$ of $\Gamma$, any functions $f,g: F \ra \{0,1\}$ and
    any matchings $\mu_1\in \mu(F,f)$ and $\mu_1\in \mu(F,g)$,
    $\sgn{\mu_1} = \sgn{\mu_2}$.
  \end{lemma}
  \begin{proof}
    It suffices to show the lemma for functions $f$ and $g$ which differ at
    exactly one edge $e \in F$ as the result then follows by transitivity.
    Moreover, by Lemma~\ref{lem:sign}, it suffices to choose any $\mu_1 \in
    \mu(F,f)$ and $\mu_2 \in \mu(F,g)$ and show that they have the same sign

    So, assume $f(e) = 0$ and $g(e)=1$ and $f$ and $g$ agree on all other edges. Let $e
    = \{v,x\}$ where $v$ has neighbours $x,y,z$ and $x$ has neighbours $v,u,w$.
    Without loss of generality, assume $\{v,y\}$ and $\{x,w\}$ are in $F$ with
    $f(\{v,y\}) = f(\{x,w\}) = g(\{v,y\}) = g(\{x,w\}) = 0$. By assumption,
    $\mu_1$ includes the edge $x_0-v_0$ and we can assume further that it
    includes the edges $v_{\emptyset}-z_0$, $v_{\{x,y\}}-x_1$ in the gadget
    corresponding to $v$ and the edges $x_{\emptyset}-u_0$, $v_{\{v,w\}}-v_1$ in
    the gadget corresponding to $x$. In other words, it contains alternating
    edges along the $10$-cycle depicted in Figure~\ref{fig:two-cycle}.
    \begin{figure}[h]
      \centering{ 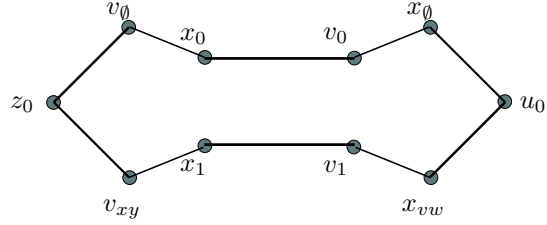
        \caption{An alternating cycle between two matchings in $\mu(F,f)$ and
          $\mu(F,g)$ where $f$ and $g$ differ on the edge $\{v,x\}$.}
        \label{fig:two-cycle}
      }
    \end{figure}
    Now, choose $\mu_2$ to be the symmetric difference between $\mu_1$ and the
    cycle in Figure~\ref{fig:two-cycle}. This is also a perfect matching and it
    is in $\mu(F,g)$ as the only difference with $\mu_1$ as far as edges between
    gadgets is concerned is that $\mu_2$ contains $x_1-v_1$ rather than
    $x_0-v_0$. Moreover, it is easily seen that $\mu_2$ is obtained from $\mu_1$
    by composition with an even permutation: namely the $5$-cycle:
    $(x_0z_0x_1x_{\{v,w\}}x_{\emptyset})$.
  \end{proof}

  With this, we are now ready to establish the main result of this section.
  \begin{lemma}\label{lem:determinant}
    If $\Gamma$ has an odd number of perfect matchings, then $\det(M) \neq 0$.
  \end{lemma}
  \begin{proof}
    By Lemma~\ref{lem:non-uniform}, we have
    \begin{eqnarray*}
      \det(M) & = & |\{ \mu \mid \mu \in \umtch(\hG) \text{ with } \sgn{\mu} = 1 \}| \\ & & - |\{ \mu \mid \mu \in \umtch(\hG) \text{ with } \sgn{\mu} = -1 \}|.
    \end{eqnarray*}
    For any $2$-factor $F$ of $\Gamma$ write $\mu(F)$ for $\bigcup_{f: F \ra
      \{0,1\}}\mu(F,f)$ and note that by Lemma~\ref{lem:fixed-Ff}
    $\vert \mu(F) \vert =  2^{4m}$ for all $F$.  Moreover, by Lemma~\ref{lem:sign-factor} all matchings
    in $\mu(F)$ have the same sign. Thus, we define $\sgn{F}$ to be $\sgn{\mu}$
    for any $\mu \in \mu(F)$. Hence, we have that
$$ \det(M) = 2^{4m}\sum_F \sgn{F},$$
where the sum is over all $2$-factors of $\Gamma$.

Since $\Gamma$ is $3$-regular, the number of $2$-factors of $\Gamma$ is exactly
the number of perfect matchings. Indeed, the complement of any $2$-factor is a
perfect matching and this gives a bijection between the collection of perfect
matchings and $2$-factors. Thus, since the number of perfect matchings of
$\Gamma$ is odd, so is the number of $2$-factors and we conclude that the sum
$\sum_F \sgn{F}$ cannot be zero, proving the result.
\end{proof}

%% file: gadget.pdf_tex
\begingroup%
  \makeatletter%
  \providecommand\color[2][]{%
    \errmessage{(Inkscape) Color is used for the text in Inkscape, but the package 'color.sty' is not loaded}%
    \renewcommand\color[2][]{}%
  }%
  \providecommand\transparent[1]{%
    \errmessage{(Inkscape) Transparency is used (non-zero) for the text in Inkscape, but the package 'transparent.sty' is not loaded}%
    \renewcommand\transparent[1]{}%
  }%
  \providecommand\rotatebox[2]{#2}%
  \newcommand*\fsize{\dimexpr\f@size pt\relax}%
  \newcommand*\lineheight[1]{\fontsize{\fsize}{#1\fsize}\selectfont}%
  \ifx\svgwidth\undefined%
    \setlength{\unitlength}{189.96297998bp}%
    \ifx\svgscale\undefined%
      \relax%
    \else%
      \setlength{\unitlength}{\unitlength * \real{\svgscale}}%
    \fi%
  \else%
    \setlength{\unitlength}{\svgwidth}%
  \fi%
  \global\let\svgwidth\undefined%
  \global\let\svgscale\undefined%
  \makeatother%
  \begin{picture}(1,0.79891969)%
    \lineheight{1}%
    \setlength\tabcolsep{0pt}%
    \put(0,0){\includegraphics[width=\unitlength,page=1]{gadget.pdf}}%
    \put(0.46750457,0.57255503){\color[rgb]{0,0,0}\makebox(0,0)[lt]{\lineheight{0}\smash{\begin{tabular}[t]{l}$x_0$\end{tabular}}}}%
    \put(0.35047412,0.57255503){\color[rgb]{0,0,0}\makebox(0,0)[lt]{\lineheight{0}\smash{\begin{tabular}[t]{l}$x_1$\end{tabular}}}}%
    \put(0.21135841,0.0785721){\color[rgb]{0,0,0}\makebox(0,0)[lt]{\lineheight{0}\smash{\begin{tabular}[t]{l}$y_1$\end{tabular}}}}%
    \put(0.10427506,0.2251237){\color[rgb]{0,0,0}\makebox(0,0)[lt]{\lineheight{0}\smash{\begin{tabular}[t]{l}$y_0$\end{tabular}}}}%
    \put(0.74021885,0.20130155){\color[rgb]{0,0,0}\makebox(0,0)[lt]{\lineheight{0}\smash{\begin{tabular}[t]{l}$z_1$\end{tabular}}}}%
    \put(0.60635339,0.07993024){\color[rgb]{0,0,0}\makebox(0,0)[lt]{\lineheight{0}\smash{\begin{tabular}[t]{l}$z_0$\end{tabular}}}}%
    \put(0.11973302,0.41304713){\color[rgb]{0,0,0}\makebox(0,0)[lt]{\lineheight{0}\smash{\begin{tabular}[t]{l}$v_{\emptyset}$\end{tabular}}}}%
    \put(0.65123756,0.41264891){\color[rgb]{0,0,0}\makebox(0,0)[lt]{\lineheight{0}\smash{\begin{tabular}[t]{l}$v_{\{y,z\}}$\end{tabular}}}}%
    \put(0,0){\includegraphics[width=\unitlength,page=2]{gadget.pdf}}%
    \put(0.48688536,0.4111968){\color[rgb]{0,0,0}\makebox(0,0)[lt]{\lineheight{0}\smash{\begin{tabular}[t]{l}$v_{\{x,z\}}$\end{tabular}}}}%
    \put(0.3122481,0.41233426){\color[rgb]{0,0,0}\makebox(0,0)[lt]{\lineheight{0}\smash{\begin{tabular}[t]{l}$v_{\{x,y\}}$\end{tabular}}}}%
    \put(0,0){\includegraphics[width=\unitlength,page=3]{gadget.pdf}}%
  \end{picture}%
\endgroup%

%% file: gadget-matching.pdf_tex
\begingroup%
  \makeatletter%
  \providecommand\color[2][]{%
    \errmessage{(Inkscape) Color is used for the text in Inkscape, but the package 'color.sty' is not loaded}%
    \renewcommand\color[2][]{}%
  }%
  \providecommand\transparent[1]{%
    \errmessage{(Inkscape) Transparency is used (non-zero) for the text in Inkscape, but the package 'transparent.sty' is not loaded}%
    \renewcommand\transparent[1]{}%
  }%
  \providecommand\rotatebox[2]{#2}%
  \newcommand*\fsize{\dimexpr\f@size pt\relax}%
  \newcommand*\lineheight[1]{\fontsize{\fsize}{#1\fsize}\selectfont}%
  \ifx\svgwidth\undefined%
    \setlength{\unitlength}{182.02829933bp}%
    \ifx\svgscale\undefined%
      \relax%
    \else%
      \setlength{\unitlength}{\unitlength * \real{\svgscale}}%
    \fi%
  \else%
    \setlength{\unitlength}{\svgwidth}%
  \fi%
  \global\let\svgwidth\undefined%
  \global\let\svgscale\undefined%
  \makeatother%
  \begin{picture}(1,0.46383209)%
    \lineheight{1}%
    \setlength\tabcolsep{0pt}%
    \put(0,0){\includegraphics[width=\unitlength,page=1]{gadget-matching.pdf}}%
    \put(-0.00311834,0.16501292){\color[rgb]{0,0,0}\makebox(0,0)[lt]{\lineheight{0}\smash{\begin{tabular}[t]{l}$v_{\emptyset}$\end{tabular}}}}%
    \put(0.15126165,0.158938){\color[rgb]{0,0,0}\makebox(0,0)[lt]{\lineheight{0}\smash{\begin{tabular}[t]{l}$z_0$\end{tabular}}}}%
    \put(0.32062061,0.15697822){\color[rgb]{0,0,0}\makebox(0,0)[lt]{\lineheight{0}\smash{\begin{tabular}[t]{l}$v_{xy}$\end{tabular}}}}%
    \put(0.61855844,0.42578426){\color[rgb]{0,0,0}\makebox(0,0)[lt]{\lineheight{0}\smash{\begin{tabular}[t]{l}$v_{xz}$\end{tabular}}}}%
    \put(0.62877737,0.00994651){\color[rgb]{0,0,0}\makebox(0,0)[lt]{\lineheight{0}\smash{\begin{tabular}[t]{l}$v_{yz}$\end{tabular}}}}%
    \put(0.85157485,0.21860457){\color[rgb]{0,0,0}\makebox(0,0)[lt]{\lineheight{0}\smash{\begin{tabular}[t]{l}$z_1$\end{tabular}}}}%
    \put(0.4680506,0.42222244){\color[rgb]{0,0,0}\makebox(0,0)[lt]{\lineheight{0}\smash{\begin{tabular}[t]{l}$x_1$\end{tabular}}}}%
    \put(0.46048547,0.01306523){\color[rgb]{0,0,0}\makebox(0,0)[lt]{\lineheight{0}\smash{\begin{tabular}[t]{l}$y_1$\end{tabular}}}}%
  \end{picture}%
\endgroup%

%% file: two-cycle.pdf_tex
\begingroup%
  \makeatletter%
  \providecommand\color[2][]{%
    \errmessage{(Inkscape) Color is used for the text in Inkscape, but the package 'color.sty' is not loaded}%
    \renewcommand\color[2][]{}%
  }%
  \providecommand\transparent[1]{%
    \errmessage{(Inkscape) Transparency is used (non-zero) for the text in Inkscape, but the package 'transparent.sty' is not loaded}%
    \renewcommand\transparent[1]{}%
  }%
  \providecommand\rotatebox[2]{#2}%
  \newcommand*\fsize{\dimexpr\f@size pt\relax}%
  \newcommand*\lineheight[1]{\fontsize{\fsize}{#1\fsize}\selectfont}%
  \ifx\svgwidth\undefined%
    \setlength{\unitlength}{219.39771652bp}%
    \ifx\svgscale\undefined%
      \relax%
    \else%
      \setlength{\unitlength}{\unitlength * \real{\svgscale}}%
    \fi%
  \else%
    \setlength{\unitlength}{\svgwidth}%
  \fi%
  \global\let\svgwidth\undefined%
  \global\let\svgscale\undefined%
  \makeatother%
  \begin{picture}(1,0.37827639)%
    \lineheight{1}%
    \setlength\tabcolsep{0pt}%
    \put(0,0){\includegraphics[width=\unitlength,page=1]{two-cycle.pdf}}%
    \put(-0.00310465,0.18445069){\color[rgb]{0,0,0}\makebox(0,0)[lt]{\lineheight{0}\smash{\begin{tabular}[t]{l}$z_0$\end{tabular}}}}%
    \put(0.16252597,0.34670915){\color[rgb]{0,0,0}\makebox(0,0)[lt]{\lineheight{0}\smash{\begin{tabular}[t]{l}$v_{\emptyset}$\end{tabular}}}}%
    \put(0.15698724,0.00825235){\color[rgb]{0,0,0}\makebox(0,0)[lt]{\lineheight{0}\smash{\begin{tabular}[t]{l}$v_{xy}$\end{tabular}}}}%
    \put(0.28901011,0.07608877){\color[rgb]{0,0,0}\makebox(0,0)[lt]{\lineheight{0}\smash{\begin{tabular}[t]{l}$x_1$\end{tabular}}}}%
    \put(0.28901011,0.30170641){\color[rgb]{0,0,0}\makebox(0,0)[lt]{\lineheight{0}\smash{\begin{tabular}[t]{l}$x_0$\end{tabular}}}}%
    \put(0.53524616,0.30321374){\color[rgb]{0,0,0}\makebox(0,0)[lt]{\lineheight{0}\smash{\begin{tabular}[t]{l}$v_0$\end{tabular}}}}%
    \put(0.53524616,0.07598106){\color[rgb]{0,0,0}\makebox(0,0)[lt]{\lineheight{0}\smash{\begin{tabular}[t]{l}$v_1$\end{tabular}}}}%
    \put(0.6785234,0.34670915){\color[rgb]{0,0,0}\makebox(0,0)[lt]{\lineheight{0}\smash{\begin{tabular}[t]{l}$x_{\emptyset}$\end{tabular}}}}%
    \put(0.66996998,0.00825235){\color[rgb]{0,0,0}\makebox(0,0)[lt]{\lineheight{0}\smash{\begin{tabular}[t]{l}$x_{vw}$\end{tabular}}}}%
    \put(0.87244911,0.18445069){\color[rgb]{0,0,0}\makebox(0,0)[lt]{\lineheight{0}\smash{\begin{tabular}[t]{l}$u_0$\end{tabular}}}}%
  \end{picture}%
\endgroup%

%% file: base-graph.tex
We have seen that if $\Gamma$ has an odd number of perfect matchings,
then the matrices $M$ and $N$ have different determinant.  In order to
play the bijection game on $M$ and $N$ we also need $\Gamma$ to be
well connected.  We now show that we can find suitable graphs that
satisfy both of these conditions simultaneously.

For a positive integer $k$, say that $\Gamma$ is \emph{$k$-well-connected} if any balanced separator of $\Gamma$ has size greater than $k$.  For our construction, we need $3$-regular bipartite graphs on $2n$ vertices which are $k$-well-connected for $k = \Omega(n)$ and which have an odd number of perfect matchings.  The main purpose of this section is to prove the existence of such a family of graphs.

\begin{theorem}\label{thm:base-graph}
  For all positive integers $n$ there is a bipartite graph $\Gamma_n = (U,V, E)$
  satisfying the following conditions:
  \begin{enumerate}
  \item $|U| = |V| =n$;
  \item $\Gamma_n$ is $3$-regular;
  \item $\Gamma_n$ is $k$-well-connected for $k = \Omega(n)$; and
  \item $\Gamma_n$ has an odd number of perfect matchings.
  \end{enumerate}
\end{theorem}

We prove the existence of these graphs by a randomized construction.  To be precise, a random $3$-regular bipartite graph on $2n$ vertices satisfies the third condition with high probability.  We show that it can be modified to satisfy the fourth condition while keeping the connectivity high.  To do this, we need some facts about the distribution of random $3$-regular bipartite graphs.

Fix $U$ and $V$ to be two disjoint sets of $n$ vertices, and we are interested in the uniform distribution on $3$-regular bipartite graphs on the vertices $U$ and $V$.  This distribution is not easy to sample from but it is known to be well-approximated by a number of other random models, including  the union of disjoint random matchings, which we now describe.  We say that a pair of bijections $\pi,\sigma: U \ra V$ is \emph{disjoint} if there is no $u \in U$ with $\pi(u) = \sigma(u)$.  Now consider a random graph $\mathcal{G}$ obtained by the following process:
\begin{enumerate}
\item choose, uniformly at random, three bijections $\pi_1,\pi_2,\pi_3: U \ra V$;
\item if for some $j \in \{1,2,3\}$ with $i\neq j$, $\pi_i$ and $\pi_j$ are not disjoint discard this choice of bijections; otherwise
\item let $\mathcal{G}$ be the bipartite graph with parts $U$ and $V$ edges $\{\{u,\pi_i(u)\} \mid i \in \{1,2,3\}\}$.
\end{enumerate}
The random graph model obtained in this way is known to be \emph{contiguous} to the uniform distribution on $3$-regular bipartite graphs~\cite{MolloyRRW97}.  This means that any property that holds asymptotically almost surely in one also holds so in the other.  The property we are interested in is that of being an expander.  It is known~\cite{BritoDH21} that a random $3$-regular bipartite graph is an expander with probability tending to $1$.
This result is, in fact, proved in the configuration model of Bollobas~\cite{Bollobas80} but this is also known to be contiguous to the uniform distribution.
We can therefore conclude that the same is true for the random graph $\mathcal{G}$.
\begin{lemma}\label{lem:expander}
  There is a constant $\alpha > 0$ such that with probability tending to $1$, $\mathcal{G}$ is an $\alpha$-expander.
\end{lemma}
An immediate consequence of this is that with high probability, $\mathcal{G}$ is $\epsilon n$-well-connected for some constant $\epsilon$.
 We now describe how we can obtain from $\mathcal{G}$ a graph which also has an odd number of perfect matchings.

Let $B_\Gamma$ denote the biadjacency matrix of $\Gamma = (U,V,E)$ with rows
indexed by $U$ and columns by $V$. Then the permanent of $B_\Gamma$ over a field
of characteristic $p$ is exactly the number of perfect matchings in $\Gamma$
modulo $p$. In particular, when $p=2$, since the permanent is the same as the
determinant, we have that the number of perfect matchings in $\Gamma$ is odd if,
and only if, $\det(B_\Gamma) \neq 0$, where the determinant is over $\ff_2$.
We do not expect that $\det(B_{\Gcal}) \neq 0$ with high probability.  To prove Theorem~\ref{thm:base-graph} it would suffice to show that this is the case with positive probability and this does seem likely to be true.  However, we adopt an indirect approach.  We show that with probability bounded away from
zero $\rank(B_\Gcal)$ is at least $n-o(n)$. And, we then show that we can transform
any graph $\Gamma$ with $\rank(B_\Gamma) < n$ to a graph $\Gamma'$ so that
$\rank(B_{\Gamma'}) > \rank(B_\Gamma) + 1$ and $\Gamma'$ is still well-connected if $\Gamma$ is. Together these give us the graphs we want. We next introduce some notation
and terminology that is helpful in establishing these two facts.

In what follows, we treat the biadjacency matrix $B_\Gamma$ of a graph $\Gamma$
as being a matrix over $\ff_2$ and so all arithmetic operations on elements of
the matrix should be taken as being over this field.

\begin{lemma}\label{lem:random-rank}
  There is a constant $\epsilon$ such that for all sufficiently large $n$,
  $\Pr[\rank(B_{\mathcal{G}}) \geq n-\epsilon \log n] \geq 1/2$.
\end{lemma}
\begin{proof}
  For any matrix $A$, if $\rank(A) = n-c$, then the dimension of the null space of $A$ is $c$, so
  there are $2^c-1$ non-zero vectors $\bd{x}$ such that $A^T\bd{x} = 0$. We show
  that for the random graph $\Gcal$, the expected number of vectors
  $\bd{x} \in \ff_2$ such that $B_{\mathcal{G}}^T\bd{x} = 0$ is at most linear in $n$\footnote{This is a very loose upper bound that suffices for our purposes.
   By more careful analysis, we can easily show that this expectation is
   bounded by a constant. Numerical simulations suggests that it in fact tends to $1$ as $n$ goes to infinity.}. Let
  $\bd{X}$ denote the random variable that is the number of such vectors.

  For an element $u \in U$, write $r_u$ for the row of $B_{\mathcal{G}}$ indexed by
  $u$. For a vector $\bd{x} \in \ff_2^U$, let $R_{\bd{x}}$ be the set $\{ u \in
  U \mid \bd{x}_u = 1 \}$ and note that the condition $B_{\mathcal{G}}^T\bd{x} = 0$ is
  equivalent to the statement $\sum_{u \in R_{\bd{x}}} r_u = 0$.

  For any set $S \subseteq U$ and a bijection $\pi: U \ra V$, we write $\pi(S)$
  to denote the image of $S$ under $\pi$. Say that $S$ is a \emph{zero-sum set}
  if $S$ is non-empty with $\sum_{u \in S} r_u = 0$. If $S$ is a zero-sum set,
  it must be the case that any vertex $v$ in $\pi_1(S) \cup \pi_2(S) \cup
  \pi_3(S)$ has exactly two neighbours in $S$. If $|S| = m$, this implies that
  $|\pi_1(S) \cap \pi_2(S)| = |\pi_2(S) \cap \pi_3(S)| = |\pi_1(S) \cap
  \pi_3(S)| = m/2$. In particular $m$ is even, and $m \leq 2n/3$.

We now estimate the expected number of zero-sum sets of size $m = 2l$ in the
random graph $\Gcal$. Fix a set $S \subseteq U$ of size $m$ and choose three permutations $\pi_1, \pi_2, \pi_3$ of $[n]$ independently at random.  Note that a
random choice of $\pi_1$ means that $\pi_1(S)$ is a uniformly random subset of
$V$ of size $m$. Thus, the probability that $|\pi_1(S) \cap \pi_2(S)| = m/2 = l$
is:
 $$ p(n,l) :=   \frac{{{2l} \choose l}{{n-2l} \choose l}}{{n \choose {2l}}}.$$
 For $S$ to be a zero-sum set, we further require that $\pi_3(S) = \pi_1(S)
 \symmdiff \pi_2(S)$. The probability that a randomly chosen $\pi_3$ gives
 exactly this set is $1/{n \choose {2l}}$. Summing over the ${n \choose {2l}}$
 choices of the set $S$, we get that the expected number of sets of 
 size $m=2l$ satisfying the condition  $|\pi_1(S) \cap \pi_2(S)| = |\pi_2(S) \cap \pi_3(S)| = |\pi_1(S) \cap
  \pi_3(S)| = m/2$, taken over all choices of  $\pi_1, \pi_2, \pi_3$
  is $p(n,l)$.  Now, the probability that two random permutations are not disjoint is the same as the probability that a random permutation contains a fixed-point, which is well known to tend to $1/e$ from above as $n$ goes to infinity.  Hence we see that the probability that all three permutations are disjoint is at least $1/e^3$ and therefore the expected number of zero-sum sets in $\mathcal{G}$ of size $2l$ is at most $e^3p(n,l)$.

  Hence the total expectation of the number of zero-sum
 sets is
 $$ \text{E}[\bd{X}]  = \sum_{1 \leq l \leq n/3} e^3p(n,l).$$
 Since $p(n,l) < 1$ for all $n$ and $l$, we get $\text{E}[\bd{X}] <
e^3 n/3$.  By Markov's inequality, it
 follows that the probability that $\bd{X} $ exceeds $2e^3 n/3$ is less than $1/2$.
 Since the dimension of the null space of $B_\Gamma$ is $\log(\bd{X} + 1)$, the
 theorem follows.
\end{proof}

To complete the construction, we show that if $\Gamma$ is a $3$-regular
bipartite graph with $n \times n$ biadjacency matrix $B_\Gamma$ and $\rank(B) <
n$, then under mild assumptions satisfied by almost all such graphs, we can edit
$\Gamma$ to get $\Gamma'$ so that $\rank(B_{\Gamma'}) > \rank(B_\Gamma)$ and
$\Gamma'$ is at least $(k-4)$-well-connected if $\Gamma$ is $k$-well-connected. 

Assume then, that $\Gamma$ is a $3$-regular graph on two sets $U$ and $V$ of $n$
vertices each and $B$ is its biadjacency matrix. As before, we write $r_u^B$ to
denote the row of $B$ indexed by $u \in U$. We drop the superscript $B$ where it
is clear from context. We always treat these rows as vectors in $\ff_2^V$.

We say that a pair of edges $e_1 = \{u_1,v_1\}$ and $e_2 = \{u_2,v_2\}$ of
$\Gamma$ with $u_1,u_2 \in U$ and $v_1,v_2 \in V$ are \emph{switchable} if they
are disjoint and neither of $\{u_2,v_1\}$ nor $\{u_1,v_2\}$ is an edge. For a
switchable pair $e_1,e_2$ we denote by $\tilde{\Gamma}_{e_1,e_2}$ the graph
obtained from $\Gamma$ by exchanging the two edges $e_1$ and $e_2$. That is,
$\tilde{\Gamma}_{e_1,e_2}$ is the bipartite graph on the vertices $U$, $V$ with
edge set
$$ E(\Gamma) \setminus \{e_1,e_2\} \cup \{ \{u_1,v_2\}, \{u_2,v_1\} \}.$$
Note that $\tilde{\Gamma}_{e_1,e_2}$ is also a $3$-regular bipartite graph. We
write $\tilde{B}_{e_1,e_2}$ for the biadjacency matrix of
$\tilde{\Gamma}_{e_1,e_2}$.

Assume now that $\rank(B) < n$. Then $B$ has a \emph{zero-sum set},
i.e.\ a set
$S \subseteq U$ such that $\sum_{u \in S} r_u = 0$. Moreover, by the
$3$-regularity of $\Gamma$, we have $2 \leq |S| \leq 2n/3$. We can now state the
lemma we aim to prove.
\begin{lemma}\label{lem:rank}
  If $B$ has a zero-sum set $S$ with $|S| < 2n/3$, then there are switchable
  edges $e_1,e_2 \in E(\Gamma)$ so that $\rank(\tilde{B}_{e_1,e_2}) > \rank(B)$.
\end{lemma}
\begin{proof}
  Fix a zero-sum set $S$ in $B$ with $|S| < 2n/3$. Let $N(S) \subseteq V$ denote
  the set of elements of $V$ which are neighbours in $\Gamma$ of vertices in
  $S$. The assumption on the size of $S$ implies that $|N(S)| < n$ and so $V
  \setminus N(S) \neq \emptyset$.

  For $i,j \in V$, write $t_{ij}$ for the vector in $\ff_2^V$ which is $1$ at
  positions $i$ and $j$ and $0$ everywhere else. Now consider the following set
  of vectors in $\ff_2^V$:
$$ T = \{ t_{ij} \mid i \in N(S) \text{ and } j \in V \setminus N(S)\}.$$

Observe that $\spn(T) =E$ where $E$ is the subspace of $\ff_2^V$ consisting of
vectors with even Hamming weight. To see this, first observe that $t_{ij} \in
\spn(T)$ for all pairs $i,j \in V$. When $i \in N(S)$ and $j \not\in N(S)$, this
is true by definition of $T$. For $i,j \in N(S)$, choose any $k \not\in N(S)$
and note that $t_{ik} + t_{jk} = t_{ij}$. Similarly, if $i,j \not \in N(S)$, we
can pick a $k \in N(S)$ and again $t_{ki} + t_{kj} = t_{ij}$. Since the
collection of vectors $\{t_{ij} \mid i,j \in V\}$ clearly spans $E$, we are
done.

Let $R = \{ r_u \mid u \in U\}$ be the set of rows of $B$. Since each $r_u$ has
Hamming weight $3$, $r_u \not\in E$ so $\spn(R) \not\subseteq E$. Since $\dim(E)
= n-1 \geq \rank(B) = \dim(\spn(R))$, we conclude that $T \not\subseteq
\spn(R)$. Let us fix a $t_{ij} \in T$ such that $t_{ij} \not\in \spn(R)$.

Since $i \in N(S)$ and $S$ is a zero-sum set, $i$ has exactly two neighbours in
$S$ and one in $U\setminus S$. On the other hand, $j$ has three neighbours, all
in $U\setminus S$. Thus, we can choose a $k \in S$ which is a neighbour of $i$
but not $j$ and an $l \in U \setminus S$ which is a neighbour of $j$ but not of
$i$. Let $e_1 = \{i,k\}$ and $e_2=\{j,l\}$ and observe that this pair is
switchable by construction.

To prove the lemma, it then suffices to prove that $\rank(\tilde{B}_{e_1,e_2}) >
\rank(B)$. We do this by establishing the following two facts.
\begin{enumerate}
\item $\sum_{u \in S}r_u^{\tilde{B}_{e_1,e_2}} \neq 0$; and
\item for any $S' \subseteq U$, if $\sum_{u \in S'}r_u^{\tilde{B}_{e_1,e_2}} =
  0$, then $\sum_{u \in S'}r_u^{B} = 0$.
\end{enumerate}
Together these establish that the null space of $\tilde{B}_{e_1,e_2}$ is
strictly smaller than that of $B$ and hence the claim.

Note that for all $u \in U \setminus \{k,l\}$, we have
$r_u^{\tilde{B}_{e_1,e_2}} = r_u^B$, while $r_k^{\tilde{B}_{e_1,e_2}} = r_k^B +
t_{ij}$ and $r_l^{\tilde{B}_{e_1,e_2}} = r_l^B + t_{ij}$.

Thus, to prove the first fact, just note that since $k$ is in $S$ and $l$ is
not, $\sum_{u \in S}r_u^{\tilde{B}_{e_1,e_2}} = \sum_{u \in S}r_u^{B} + t_{ij} =
t_{ij}$.

To prove the second fact, consider any set $S' \subseteq U$ such that $\sum_{u
  \in S'}r_u^{\tilde{B}_{e_1,e_2}} = 0$. We consider the following cases.
\begin{itemize}
\item If neither $k$ nor $l$ is in $S'$, then $\sum_{u \in
    S'}r_u^{\tilde{B}_{e_1,e_2}} = \sum_{u \in S'}r_u^B$, so since the former
  sum is $0$, so is the latter.
\item If both $k$ and $l$ are in $S'$, then $\sum_{u \in
    S'}r_u^{\tilde{B}_{e_1,e_2}} = \sum_{u \in S'}r_u^B + 2t_{ij} = \sum_{u \in
    S'}r_u^B$, and the same argument applies.
\item If exactly one of $k$ and $l$ is in $S'$, then $\sum_{u \in
    S'}r_u^{\tilde{B}_{e_1,e_2}} = \sum_{u \in S'}r_u^B + t_{ij}$. Hence, if
  $\sum_{u \in S'}r_u^{\tilde{B}_{e_1,e_2}} = 0$, we must have $\sum_{u \in
    S'}r_u^B = t_{ij}$. However, $i$ and $j$ were chosen so that $t_{ij} \not\in
  \spn(R)$, so this is impossible.
\end{itemize}
\end{proof}

\begin{proof}[Proof of Theorem~\ref{thm:base-graph}]
  By Lemma~\ref{lem:expander}, for large enough values of $n$, the random $3$-regular graph $\mathcal{G}$ is $\tau n$-well-connected for some
  constant $\tau > 0$ with probability tending to $1$. Thus, with 
  high probability, the first three conditions are satisfied. If the biadjacency
  matrix $B_{\Gamma}$ of the resulting graph $\Gamma$ has rank $n$, we are done.

  If $\rank(B_\Gamma) < n$, then with probability at least $1/2$, we have
  $\rank(B_\Gamma) \geq n - \epsilon \log n$ by Lemma~\ref{lem:random-rank}.
  Moreover, since the expected number of zero-sum sets of size exactly $2n/3$ is
  at most  $e^3 p(n,n/3)$ by the argument in the proof of
  Lemma~\ref{lem:random-rank}, and this value tends to $0$ as $n$ grows, with
  high probability, $B_{\Gamma}$ has no zero-sum sets of this size. Hence, with
  positive probability, $\Gcal$ satisfies
  the pre-conditions of Lemma~\ref{lem:rank}.

  Note that if $\Gamma$ satisfies the conditions of Lemma~\ref{lem:rank}, then
  any zero-sum set in the graph $\tilde{\Gamma}_{e_1,e_2}$ is also a zero-sum
  set in $\Gamma$. Hence, if $\Gamma$ contains no zero-sum sets of size exactly
  $2n/3$, the same is true of $\tilde{\Gamma}_{e_1,e_2}$. We can thus repeatedly
  apply the construction of Lemma~\ref{lem:rank} to obtain a graph $\Gamma'$
  such that $\rank(B_\Gamma) = n$. It then follows that $\Gamma'$ has an odd
  number of perfect matchings. It remains to argue that 
  $\Gamma'$ is still well-connected. Note that since $\rank(B_\Gamma) \geq n -
  \epsilon \log n$ , we get $\Gamma'$ from $\Gamma$ by at most $\epsilon \log n$
  applications of Lemma~\ref{lem:rank}. At each step edges incident at most $4$
  vertices are modified.  Thus, if $S$ is a balanced separator in $\Gamma'$, then adding these four vertices to it gives us a balanced separator in $\Gamma$.  Thus, if $\Gamma$ is $k$-well-connected, then $\Gamma'$ is at least $(k-4)$-well-connected  and the result is  proved.
  
\end{proof}

%% file: playing.tex
Suppose $\Gamma$ is a bipartite $3$-regular graph on two sets $U$ and $V$ of $m$
vertices each that is $(k+3)$-well-connected. Let $\hG$ be the CFI-graph constructed
from $\Gamma$ as described in Section~\ref{sec:graph}, and $M$ and $N$ be two
biadjacency matrices for $\hG$ where $N$ is obtained from $M$ by interchanging
exactly one pair of columns. We aim to prove that Duplicator has a winning
strategy in the $(\alt_n \times \alt_n,k)$-bijection game played on $M$
and $N$.

Recall that $\hG$ is a bipartite graph on two sets $X$ and $Y$ of $n = 10m$
vertices each with $X = \bigcup_{u\in U} I_u \cup \bigcup_{v\in V} O_v$ and $Y =
\bigcup_{v\in V} I_u \cup \bigcup_{u\in U} O_u$.  
To say that
Duplicator has a winning strategy in the $(\alt_n \times\alt_n,k)$-bijection game played on $M$ and $N$ is the same as saying that
Duplicator has a winning strategy in the $(\alt_n
\times\alt_n,k)$-bijection game played on the pair of graphs $\hG$ and
$\hG'$ where the latter is obtained from $\hG$ by swapping two
elements $y,y' \in Y$.  It does
not matter which two elements $y,y'$ we choose as any swap can be obtained from
$(yy')$ by composing with a permutation in $\alt_Y$.  So, for some $u \in U$, fix two vertices $x_0,x_1 \in O_u$
which form a single pair in the gadget corresponding to $u$ and let $\alpha =
(x_0x_1)$.  We write $\alpha\hG$ for the graph on $X \cup Y$ which is
exactly the same as $\hG$ except the neighbours of $x_0$ in
$\alpha\hG$ are exactly the neighbours of $x_1$ in $\hG$ and the neighbours of $x_1$ in
$\alpha\hG$ are exactly the neighbours of $x_0$ in $\hG$.

\begin{lemma}\label{lem:winning}
  Duplicator has a winning strategy in the $(\alt_X \times
  \alt_Y,k)$-bijection game played on the graphs $\hG$ and $\alpha\hG$.
\end{lemma}

To prove this lemma, we first introduce some notation and some observations
about $\hG$. Consider permutations of the vertices $I_v \cup O_v$ in the gadget
in Figure~\ref{fig:gadget} which fix each of the sets $\{x_0,x_1\}$,
$\{y_0,y_1\}$ and $\{z_0,z_1\}$ setwise. It is easily checked (and this is the
key property of the gadget) that for any two of these three sets there is an
autormorphism of the gadget that swaps the two vertices inside the two sets while leaving the two vertices in the third set fixed.  For
instance, there is an automorphism which we denote $\beta^v_{xy}$, that
exchanges $x_0$ with $x_1$ and $y_0$ with $y_1$. The action of this automorphism
on $I_v$ is to swap $v_{\emptyset}$ with $v_{\{x,y\}}$ and $v_{\{x,z\}}$ with
$v_{\{y,z\}}$. Each such automorphism consists of two swaps in $I_v$ and two in
$O_v$ and so is a permutation in $\alt_X \times \alt_Y$.

We can compose such automorphisms of the individual gadgets to get certain
automorphisms of $\hG$. Let $C = v_1\cdots v_l$ be a simple cycle in the graph
$\Gamma$. That is, there are edges from $v_1$ to $v_{i+1}$ for each $i$ with
$1\leq i < l$ and an edge from $v_l$ to $v_1$. We define the permutation
$\beta_C$ of $X \cup Y$ as the
composition $$\beta^{v_1}_{v_lv_2}\beta^{v_2}_{v_1v_3}\cdots\beta^{v_i}_{v_{i-1}v_{i+1}}\cdots\beta^{v_l}_{v_{l-1}v_1}.$$
This is easily seen to be an automorphism of $\hG$.  Since it is the compostion
of permutations each of which is in $\alt_X \times \alt_Y$,  $\beta_C
\in \alt_X \times \alt_Y$.

Now, say that a permutation $\beta$ of $X \cup Y$ is \emph{coherent} if for each
$v \in U \cup V$, $\beta(I_v) = I_v$ and $\beta(O_v) = O_v$. We are only
interested in coherent permutations.  Let vertices $u \in U$ and $v \in V$ be
neighbours in $\Gamma$ and let $v_0,v_1$ be the pair of vertices in $O_u$ which
connect to vertices in $O_v$. We say that a coherent permutation $\beta$ is
\emph{good bar $uv$} if composing it with the swap $(v_0v_1)$ yields an
automorphism of $\hG$. We are now ready to describe the Duplicator winning
strategy.
\begin{proof}[Proof of Lemma~\ref{lem:winning}]
  We describe Duplicator's winning strategy. The position at any point of the
  game with Spoiler to move, consists of up to $k$ pebbled vertices of $\hG$
  along with a permutation $\beta$ of $X \cup Y$ which is obtained from the
  initial permutation $\alpha$ by means of composing it with an element of
  $\alt_X \times \alt_Y$.  If $x$ is the vertex covered by pebble $i$, let $p_i \in U \cup V$ be the vertex of $\Gamma$ such that $x \in I_{p_i} \cup O_{p_i}$.  In other words $p_1,\ldots,p_k$ enumerate the vertices of $\Gamma$ whose gadgets in $\hG$ contain the pebbled vertices.  Note that by the connectedness assumption, $\Gamma\setminus\{p_1,\ldots,p_k\}$ has a component $\Delta$ which contains more than half of the vertices of $\Gamma$, and $\Delta$ is $2$-connected.  We call $\Delta$ the \emph{large} component at this game position.

  We show that Duplicator can play to  maintain the following invariant:
  \begin{description}
  \item[(*)] there is an  edge
  $\{u,v\}$ of $\Gamma$ such that $u \in U$ and $v \in V$ are both in the large component and $\beta$ is a coherent permutation that is good bar $uv$.
  \end{description}
  In particular, this guarantees that $\beta$ is a partial
  isomorphism on the pebbled positions.  Indeed, $\beta$ is a partial isomorphism on the graph $\hG$ excluding $I_u \cup O_u \cup I_v \cup O_v$, and none of these vertices is pebbled.
Thus, if Duplicator can maintian the
  invariant (*) it is a winning strategy.

  It is clear that the initial bijection $\alpha$ satisfies (*) as no vertices are pebbled.

  At each subsequent move, Spoiler places a pebble on a vertex $x$.
Duplicator  chooses any edge $\{u',v'\}$ in the large component (this could be $\{u,v\}$ if they are still in the large component).
  We then distinguish two cases.
   \begin{enumerate}
  \item If $x \not\in \{v_0,v_1\}$ then Duplicator's response is
  to compose $\beta$ with $(v_0v_1)(v_0'v_1')$.  This is a valid move as none of
  the four vertices is pebbled and the pemutation $(v_0v_1)(v_0'v_1')$ is in
  $\alt_X \times \alt_Y$ since all four of $v_0,v_1,v_0',v_1'$ are in $Y$.  Moreover, the fact that $\beta$ is good bar $uv$ implies that composing it with
    $(v_0v_1)$ yields an automorphism of $\hG$. Composing this automorphism with
    $(v_0'v_1')$ gives us a permutation that is good bar $u'v'$.  Thus, the invariant (*) is maintained.
  \item If $x \in \{v_0,v_1\}$ Duplicator must compose $\beta$ with a
    permutation that fixes $x$, so in fact fixes both $v_0$ and $v_1$. By the fact that the large component $\Delta$ before the pebble is placed on $x$ is $3$-connected, we know that  there is a path in $\Delta$ from $u$ to $v$
    that does not use the edge $e = \{u,v\}$. Combining this with $e$ we obtain
    a simple cycle $C$. Then, $\beta_C$ is an automorphism of $\hG$ which, in
    particular, swaps $v_0$ and $v_1$. Duplicator's move is to compose $\beta$
    with the permutation $(v_0,v_1)\beta_C(v_0',v_1')$. Call the resulting
    permutation $\beta'$. Observe first that this is a valid move, as $\beta_C
    \in \alt_X \times \alt_Y$ and we are composing it with two swaps of elements
    of $Y$. It remains to argue that $\beta'$ is good bar $u'v'$. By assumption
    composing $\beta$ with $(v_0v_1)$ yields an automorphism of $\hG$ and
    $\beta_C$ is also an automorphism of $\hG$. Thus, $\beta(v_0,v_1)\beta_C$ is
    an automorphism of $\hG$ and $\beta'$ is just the composition of this with
    $(v_0',v_1')$.
  \end{enumerate}
\end{proof}

%% file: main-proof.tex
We pull things together to prove Theorem~\ref{thm:main}

\begin{proof}[Proof of Theorem~\ref{thm:main}]
  We have from Theorem~\ref{thm:base-graph} that for each $n \in \nats$ there
  exists a
  $3$-regular balanced bipartite graph $\Gamma_n$ with $2n$ vertices that is $k(n)$-well-connected for $k(n) = \Omega(n)$ and has an odd number of perfect matchings. Let $\hG_n$ be the CFI-graph constructed
from $\Gamma_n$ as described in Section~\ref{sec:graph}, and $M_n$ and $N_n$ be two
biadjacency matrices for $\hG_n$ where $N_n$ is obtained from $M_n$ by interchanging
exactly one pair of columns. From Lemma~\ref{lem:winning} we have that Duplicator
has a winning strategy for the $(\alt_n \times \alt_n, k(n) - 3)$-bijection game on $M_n$ and $N_n$. From Lemma~\ref{lem:determinant}
  and the fact that $\Gamma_n$ has an odd number of perfect matchings, it
  follows that $\det(M_n) \neq 0$ and so, since $\det(M_n) = -\det(N_n)$,  we
  have $\det(M_n) \neq \det(N_n)$. The result now follows from
  Theorem~\ref{thm:general-lower-bound}. 
\end{proof}

Theorem~\ref{thm:main} is stated and proved specifically for fields of characteristic zero.
 We could prove the result for fields of characteristic $p$, when $p$ is an odd prime, provided that the matrices $M_n$ we construct have non-zero determinant modulo $p$.  As noted in the proof of Lemma~\ref{lem:determinant}, $\det(M_n) = 2^{4m} \sum_F\sgn{F}$, where the sum is over all $2$-factors (or equivalently over all perfect matchings) of the bipartite graph $\Gamma_n$.  Of course, $\sum_F\sgn{F}$ is just the determinant of the bi-adjacency matrix $B_{\Gamma_n}$ of $\Gamma_n$.  Hence, for odd prime $p$, $\det{M_n} \not\equiv 0 \pmod p$ provided that $\det(B_{\Gamma_n}) \not\equiv \pmod p$.  Now, the construction in Section~\ref{sec:base-graph} is aimed at ensuring that $\det(B_{\Gamma_n}) \not\equiv 0 \pmod 2$.  It seems plausible that we could just as well ensure that $\det(B_{\Gamma_n}) \not\equiv 0 \pmod p$ for some odd prime $p$.  This would immediately yield the analogue of Theorem~\ref{thm:main} in characteristic $p$.  We leave this extension to future work.

%% file: permanent.tex
We previously established in~\cite{DawarW20} lower bounds on  symmetric circuits for
the permanent showing that there are no subexponential square-symmetric
circuits computing the permanent of an $n \times n$ matrix in any field of
characteristic zero, along with a similar result for matrix-symmetric circuits in any field of characteristic other than two.

The two bounds are consequences of the same construction: we give, for each $k$,
a pair of bipartite graphs $X_k$ and $\tilde{X}_k$ on which Duplicator has a
winning strategy in the $k$-pebble bijection game and which have different numbers
of perfect matchings.  The graphs $X_k$ and $\tilde{X}_k$ are on two sets $A$
and $B$ of $n = O(k)$ vertices each and the difference between the number of
perfect matchings in $X_k$ and $\tilde{X}_k$ is a power of $2$.  The $k$-pebble
bijection game for which a Duplicator winning strategy is shown is essentially
the $(\sym_A \times \sym_B,k)$ game.  This shows that the biadjacency
matrices of $X_k$ and $\tilde{X}_k$ cannot be distinguished by $\sym_A \times 
\sym_B$-symmetric circuits of subexponential size and also that the adjacency
matrices of $X_k$ and $\tilde{X}_k$ cannot be distinguished by $\sym_{A\cup
  B}$-symmetric circuits of subexponential size.  Since $X_k$ and $\tilde{X}_k$
have different numbers of perfect matchings, their biadjacency matrices $B_{X_k}$ and $B_{\tilde{X}_k}$ have
distinct permanents.  Since the number of perfect matchings differ by a power of
$2$, they have distinct permanents modulo $p$ for any odd prime $p$.  Moreover,
since the permanent of the adjacency matrix of a bipartite graph is the square
of the permanent of its biadjacency matrix, we have that the adjacency matrices $A_{X_k}$ and $A_{\tilde{X}_k}$ also
have distinct permanents.  Together, these give us the stated lower bounds for
circuits computing the permanent in the second and fourth columns of
Table~\ref{tab:results}.  To establish the lower bound in the third column, it
suffices to observe that Duplicator has a winning strategy on $X_k$ and
$\tilde{X}_k$ even in the restricted game $(\alt_A \times \alt_B,k)$.  To see this, we give a brief account of the construction, and it is instructive to contrast it with the graphs we use in Section~\ref{sec:graph}.

\begin{figure}[h]
  \centering{ 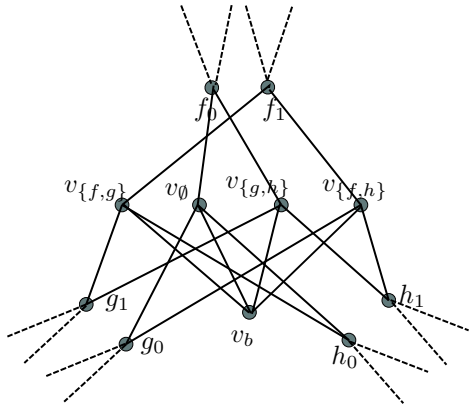
    \caption{A gadget in $X(\Gamma)$ corresponding to vertex $v$ with incident edges
      $f,g,h$}
    \label{fig:permanent}
  }
\end{figure}

The construction of the graph $\hG$ from $\Gamma$ given in Section~\ref{sec:graph} is essentially the original construction given by Cai et al.~\cite{CFI92}.  Their construction gives from $\Gamma$ a pair of non-isomorphic graphs which are not distinguishable by $k$-dimensional Weisfeiler-Leman equivalence.  We use only one graph $\hG$ from the pair, as our game is played on two different biadjacency matrices of the same graph and our main concern is that this matrix has non-zero determinant.  Of course, the different biadjacency matrices have the same permanent, and for a lower bound for the latter, we do need to play the game on a pair of non-isomorphic graphs.  So, we look again at the pairs of graphs given by the CFI construction.
The other graph in the pair  would be obtained from $\hG$ by ``twisting'' one of the edges.  That is, for some edge $e = \{u,v\}$ of $\Gamma$ we replace the two edges $e_0 = \{u_0,v_0\}$ and $e_1 = \{u_1,v_1\}$ by the edges $\{u_0,v_1\}$ and $\{u_1,v_0\}$.  It can be verified that the number of perfect matchings in the two graphs is the same.  Thus, the permanents of the biadjacency matrices of the two graphs are the same and they cannot be used directly to establish the lower bounds we want.  In the construction we presented in~\cite{DawarW20} we adapted the CFI construction in two ways.  The graph $X(\Gamma)$ is obtained from $\hG$ by first, for each  edge $e = \{u,v\}$ of $\Gamma$, contracting the two edges $e_0 = \{u_0,v_0\}$ and $e_1 = \{u_1,v_1\}$ in $\hG$ and secondly, for each vertex $v$ of $\Gamma$, adding a new vertex $v_b$ to $\hG$ which is adjacent to all four vertices in $I_v$.  Overall, this is equivalent to replacing each vertex $v$ of $\Gamma$ with incident edges $f,g,h$ with the gadget in Figure~\ref{fig:permanent} where the dashed lines indicate edges whose endpoints are in other gadgets.

The resulting graph $X(\Gamma)$ is a $4$-regular bipartite graph and $\tilde{X}(\Gamma)$ is obtained from it by taking one vertex $v$ of $\Gamma$ and in the corresponding gadget, for one edge $e$ incident on $v$, interchanging the connections of $e_0$ and $e_1$.  The fact that  $X(\Gamma)$  and $\tilde{X}(\Gamma)$ have biadjacency matrices with different permanents is proved in~\cite{DawarW20}.  It is interesting to note, however, that these matrices have determinant zero, so do not yield a lower bound on the determinant.

Finally, we briefly describe a winning strategy for Duplicator in the $(\alt_A\times \alt_B,k)$-bijection game played on $X(\Gamma)$ and $\tilde{X}(\Gamma)$.  The winning strategy in the ordinary $k$-pebble bijection game, is described in~\cite{DR07}
and is based on the fact that the graph $\Gamma$ has tree-width greater than $k$
and so there is a winning strategy for robber in the $k$-cops-and-robbers game
played on $\Gamma$.  This is lifted to a winning strategy for Duplicator in the
$k$-bijection game which sees Duplicator maintain a bijection $\beta$ that is an
isomorphism except at one gadget corresponding to a vertex $v$ of $\Gamma$.  The
position $v$ is given as a robber winning position in a $k$-cops-and-robbers
game played on the graph $\Gamma$.  At each move, Duplicator takes a path from
$v$ to $v'$ in $\Gamma$ describing a robber move and changes $\beta$ to a
bijection $\beta'$ by composing it with automorphisms for the gadgets
corresponding to the vertices along the path.  This results in $\beta'$ being an
isomorphism except at the gadget corresponding to $v'$.  It can be easily
checked that $\beta'$ is obtained from $\beta$ by composing with a permutation
in $\alt_A \times \alt_B$ provided that the path from $v$ to $v'$ is of even length.  It is easy to ensure that the graph $\Gamma$ of tree-width greater than $k$ used is bipartite and robber has a winning strategy in the $k$-cops and robber game in which robber always ends up on one side of the bipartition.  This gives us the desired result.
\begin{theorem}\label{thm:permanent}
    Let $\ff$ be a field of characteristic other than $2$. There is no family of
  $\alt_n \times \alt_n$-symmetric circuits $(C_n)_{n \in \nats}$ over $\ff$ of size $2^{o(n)}$  computing the permanent over $\ff$.
\end{theorem}

%% file: gadget2.pdf_tex
\begingroup%
  \makeatletter%
  \providecommand\color[2][]{%
    \errmessage{(Inkscape) Color is used for the text in Inkscape, but the package 'color.sty' is not loaded}%
    \renewcommand\color[2][]{}%
  }%
  \providecommand\transparent[1]{%
    \errmessage{(Inkscape) Transparency is used (non-zero) for the text in Inkscape, but the package 'transparent.sty' is not loaded}%
    \renewcommand\transparent[1]{}%
  }%
  \providecommand\rotatebox[2]{#2}%
  \newcommand*\fsize{\dimexpr\f@size pt\relax}%
  \newcommand*\lineheight[1]{\fontsize{\fsize}{#1\fsize}\selectfont}%
  \ifx\svgwidth\undefined%
    \setlength{\unitlength}{174.07873441bp}%
    \ifx\svgscale\undefined%
      \relax%
    \else%
      \setlength{\unitlength}{\unitlength * \real{\svgscale}}%
    \fi%
  \else%
    \setlength{\unitlength}{\svgwidth}%
  \fi%
  \global\let\svgwidth\undefined%
  \global\let\svgscale\undefined%
  \makeatother%
  \begin{picture}(1,0.86427334)%
    \lineheight{1}%
    \setlength\tabcolsep{0pt}%
    \put(0,0){\includegraphics[width=\unitlength,page=1]{gadget2.pdf}}%
    \put(0.40434076,0.62346931){\color[rgb]{0,0,0}\makebox(0,0)[lt]{\lineheight{0}\smash{\begin{tabular}[t]{l}$f_0$\end{tabular}}}}%
    \put(0.55236888,0.62346931){\color[rgb]{0,0,0}\makebox(0,0)[lt]{\lineheight{0}\smash{\begin{tabular}[t]{l}$f_1$\end{tabular}}}}%
    \put(0.21346247,0.21249671){\color[rgb]{0,0,0}\makebox(0,0)[lt]{\lineheight{0}\smash{\begin{tabular}[t]{l}$g_1$\end{tabular}}}}%
    \put(0.28942423,0.11900527){\color[rgb]{0,0,0}\makebox(0,0)[lt]{\lineheight{0}\smash{\begin{tabular}[t]{l}$g_0$\end{tabular}}}}%
    \put(0.84842498,0.21833989){\color[rgb]{0,0,0}\makebox(0,0)[lt]{\lineheight{0}\smash{\begin{tabular}[t]{l}$h_1$\end{tabular}}}}%
    \put(0.70234463,0.08589375){\color[rgb]{0,0,0}\makebox(0,0)[lt]{\lineheight{0}\smash{\begin{tabular}[t]{l}$h_0$\end{tabular}}}}%
    \put(0.34171244,0.45353846){\color[rgb]{0,0,0}\makebox(0,0)[lt]{\lineheight{0}\smash{\begin{tabular}[t]{l}$v_{\emptyset}$\end{tabular}}}}%
    \put(0.47311823,0.46825384){\color[rgb]{0,0,0}\makebox(0,0)[lt]{\lineheight{0}\smash{\begin{tabular}[t]{l}$v_{\{g,h\}}$\end{tabular}}}}%
    \put(0,0){\includegraphics[width=\unitlength,page=2]{gadget2.pdf}}%
    \put(0.68215613,0.46116013){\color[rgb]{0,0,0}\makebox(0,0)[lt]{\lineheight{0}\smash{\begin{tabular}[t]{l}$v_{\{f,h\}}$\end{tabular}}}}%
    \put(0,0){\includegraphics[width=\unitlength,page=3]{gadget2.pdf}}%
    \put(0.12385572,0.45964684){\color[rgb]{0,0,0}\makebox(0,0)[lt]{\lineheight{0}\smash{\begin{tabular}[t]{l}$v_{\{f,g\}}$\end{tabular}}}}%
    \put(0,0){\includegraphics[width=\unitlength,page=4]{gadget2.pdf}}%
    \put(0.48466549,0.13401069){\color[rgb]{0,0,0}\makebox(0,0)[lt]{\lineheight{0}\smash{\begin{tabular}[t]{l}$v_b$\end{tabular}}}}%
    \put(0,0){\includegraphics[width=\unitlength,page=5]{gadget2.pdf}}%
  \end{picture}%
\endgroup%

%% file: conclusions.tex
The study of the complexity of symmetric circuits began in the context of logic.  Specifications of decision problems on graphs (or similar structures) formulated in formal logic translate naturally into algorithms that respect the symmetries of the graphs.  This yields a restricted model of computation based on symmetric circuits for which we are able to prove concrete lower bounds, in a fashion similar to the restriction to monotone circuits.  Methods developed in the realm of logic for proving inexpressibility results can be reinterpreted as circuit lower bound results.

One step in this direction was the connection established in~\cite{AndersonD17} between polynomial-size Boolean threshold circuits on the one hand and fixed-point logic with counting on the other.  This shows that the power of symmetric Boolean threshold circuits to decide graph properties is delimited by the counting width of those properties.  In particular, this shows that a number of $\NP$-complete graph problems including $3$-colourability and Hamiltonicity cannot be decided by polynomial-size symmetric Boolean threshold circuits.  This is particularly interesting as the power of such symmetric circuits has been shown to encompass many strong algorithmic methods based on linear and semidefinite programming (see, for instance, \cite{AtseriasDO21}).  This methodology was extended to graph parameters beyond decision problems, and to arithmetic circuits rather than Boolean circuits in~\cite{DawarW20}.  Together these extensions established that no subexponential size square symmetric (i.e.\ unchanged by simultaneous row and column permutations) arithmetic circuits could compute the permanent.

The permanent of a $\{0,1\}$ matrix $M$ has a natural interpretation as a graph invariant and so lends itself easily to methods for proving lower bounds on graph parameters.  The situation with the determinant of $M$ is more subtle.  If $M$ is a symmetric $n \times n$ matrix, then we can see at as the adjacency matrix of a graph $\Gamma$ on $n$ vertices and the determinant is a graph invariant, that is to say it only depends on the isomorphism class of $\Gamma$.  Moreover, it is a graph invariant that can be computed efficiently by symmetric circuits (at least in characteristic zero) as was shown for Boolean circuits in~\cite{Holm10} and for arithmetic circuits in~\cite{DawarW20}.  When $M$ is not a symmetric matrix, we could think of it as the adjacency matrix of a directed graph.  In this case, the symmetries that a circuit must preserve are still the permutations of $n$, so simultaneous permutations of the rows and columns of $M$ and the upper bounds obtained still apply.  But, we can also think of $M$ as a biadjacency matrix of a bipartite graph $\Gamma$, and now the determinant of $M$ is not an invariant of $\Gamma$.   We have a richer set of symmetries, and methods for proving bounds on the complexity of graph parameters do not directly apply.

What we have sought to do in the present paper is to develop the methods for
proving lower bounds on the counting width of graph parameters to a general
methodology for proving circuit lower bounds for polynomials or more generally
functions invariant under certain permutations of their input variables.  To do
this, we prove a support theorem for circuits which is for a more general
collection of symmetry groups than proved in prior literature; we adapt the
Spoiler-Duplicator bijection game to work for more general invariance groups and more general structured 
inputs than those arising as symmetries of graph matrices; and we show a direct
relationship between these games and orbit size of circuits that bypasses
connections with width measures on graphs.  This methodology is then applied to
arithmetic circuits computing the determinant and we are able to prove an
exponential lower bound for circuits symmetric under the full permutation group
$D_n \leq \sym_n \times \sym_n$ that fixes the determinant of $M$.  Indeed, we
do this for the smaller group $\alt_n \times \alt_n$.  The application requires considerable work in constructing the example matrices and applying the bijection games.

We see one main contribution to be establishing the general methodology for
proving circuit lower bounds under various notions of symmetry.  There are many
ways in which this could be pushed further.   First, our proof of the support
theorem requires the presence of large alternating groups in the symmetry group
under consideration.  Perhaps more sophisticated notions of support could be
developed which would allow us to consider smaller groups.  Secondly, while we
state our main result for the group $\alt_n \times \alt_n$, the bijection game itself uses rather fewer symmetries.  It would be interesting to establish tighter bounds on the symmetry group for which we get exponential lower bounds.  Indeed, the results can be seen as giving a trade-off between circuit size and symmetries and this suggests an interesting terrain in which to explore the symmetry requirements of the circuit as a resource.

\paragraph*{Acknowledgements}  We are grateful to Benedikt Pago, whose comments resulted in a much improved version of the definitions and results in Section~\ref{sec:bij-games}.  We are also grateful to Albert Atserias for useful discussions on the construction in Section~\ref{sec:base-graph}.